\documentclass[twoside,11pt]{article}

\usepackage[accepted]{melba}

\usepackage{amsmath,amsfonts}

\usepackage[inline]{enumitem}
\usepackage{booktabs}
\usepackage{collcell}
\usepackage{xcolor}
\graphicspath{{figs/}}
\usepackage{siunitx}
\usepackage{pifont}

\newcommand{\cmark}{\ding{51}}%
\newcommand{\xmark}{\ding{55}}%

\newcolumntype{C}[1]{>{\centering\let\newline\\\arraybackslash\hspace{0pt}}m{#1}}
\newcommand{\frechet}{Fr\'{e}chet}

\newcommand{\textblue}[1]{\leavevmode\color{black}{#1}\color{black}}

\melbaheading{2022:024}{https://www.melba-journal.org/papers/2022:024.html}{2022}{1-36}{12/2021}{08/2022}{Ellis, Martinez Manzanera, Baltatzis, Nawaz, Nair, Le Folgoc, Desai, Glocker, Schnabel}{}{}

\ShortHeadings{3D GANs for Lung Tissue Modelling in Pulmonary CT}{Ellis et al}
\firstpageno{1}

\title{Evaluation of 3D GANs for Lung Tissue Modelling in Pulmonary CT}

\author{\name Sam~Ellis \email sam.ellis@kcl.ac.uk \\  
	\addr School of Biomedical Engineering and Imaging Sciences, King's College London, UK
	\AND
	\name Octavio~E.~Martinez~Manzanera \\
	\addr School of Biomedical Engineering and Imaging Sciences, King's College London, UK
	\AND
	\name Vasileios~Baltatzis \\
	\addr School of Biomedical Engineering and Imaging Sciences, King's College London, UK\\
	\addr Biomedical Image Analysis Group, Imperial College London, UK
	\AND
	\name Ibrahim~Nawaz \\
	\addr School of Biomedical Engineering and Imaging Sciences, King's College London, UK
	\AND
	\name Arjun~Nair \\
	\addr Department of Radiology, University College London, UK
	\AND
	\name Lo\"{i}c~Le~Folgoc \\
	\addr Biomedical Image Analysis Group, Imperial College London, UK
	\AND
	\name Sujal~Desai \\
	\addr The Royal Brompton \& Harefield NHS Foundation Trust, UK
	\AND
	\name Ben~Glocker \\
	\addr Biomedical Image Analysis Group, Imperial College London, UK
	\AND
	\name Julia~A.~Schnabel \\
	\addr School of Biomedical Engineering and Imaging Sciences, King's College London, UK \\
	\addr Institute of Machine Learning in Biomedical Imaging, Helmholtz Center Munich, Germany \\
	\addr Faculty of Informatics, Technical University of Munich, Germany
}

\begin{document}

\maketitle

\begin{abstract}

\textblue{Generative adversarial networks (GANs) are able to model accurately the distribution of complex, high-dimensional datasets, for example images. This characteristic makes high-quality GANs useful for unsupervised anomaly detection in medical imaging. However, differences in training datasets such as output image dimensionality and appearance of semantically meaningful features mean that GAN models from the natural image processing domain may not work `out-of-the-box' for medical imaging applications, necessitating re-implementation and re-evaluation. In this work we adapt and evaluate three GAN models to the application of modelling 3D healthy image patches for pulmonary CT. To the best of our knowledge, this is the first time that such a detailed evaluation has been performed. The deep convolutional GAN (DCGAN), styleGAN and the bigGAN architectures were selected for investigation due to their ubiquity and high performance in natural image processing. We train different variants of these methods and assess their performance using the widely used \frechet\ Inception Distance (FID). In addition, the quality of the generated images was evaluated by a human observer study, the ability of the networks to model 3D domain-specific features was investigated, and the structure of the GAN latent spaces was analysed. Results show that the 3D styleGAN approaches produce realistic-looking images with meaningful 3D structure, but suffer from mode collapse which must be explicitly addressed during training to obtain diversity in the samples. Conversely, the 3D DCGAN models show a greater capacity for image variability, but at the cost of poor-quality images. The 3D bigGAN models provide an intermediate level of image quality, but most accurately model the distribution of selected semantically meaningful features. The results suggest that future development is required to realise a 3D GAN with sufficient representational capacity for patch-based lung CT anomaly detection and we offer recommendations for future areas of research, such as experimenting with other architectures and incorporation of position-encoding.}

\end{abstract}


\section{Introduction}
\label{sec:intro}

Generative modelling of complex datasets (e.g. images) has advanced greatly in recent years with the advent and development of generative adversarial networks (GANs) \citep{Goodfellow2014}. GANs are a class of deep learning methods that in general pit two networks against each other in a `game' wherein the goal of one network (the discriminator) is to distinguish between real samples and the output from the other network (the generator), while the goal of the generator is to produce samples that are difficult for the discriminator to correctly identify as real or not. By training the two networks in an adversarial fashion, the generator can learn to produce realistic random outputs for even complex data distributions. 

GANs have primarily been developed in natural image processing due to the availability of large public datasets, and in that domain they have found particular application in image manipulation, whereby conditioning on an input image allows for non-linear image filtering, scene modification, and attribute switching \citep{Perarnau2016,Lample2017,Zhu2017,Isola2018}. 

GANs have also been widely applied in medical imaging, across a range of tasks such as image-to-image translation (including segmentation) \citep{Emami2018,Dong2019}, advanced data augmentation, \citep{Liu2018,Frid-Adar2018,Bu2021,Barile2021}, image reconstruction and restoration \citep{Reader2020,Ravishankar2020,Montalt-Tordera2021}, and anomaly detection \citep{Schlegl2019}. See \cite{Yi2019} for a recent review of GANs in medical imaging. 


These latter two problems, anomaly detection \citep{Schlegl2019} and image reconstruction/restoration \citep{Reader2020,Ravishankar2020,Montalt-Tordera2021}, serve as the motivation for this work. For anomaly detection, an accurate model of the entire distribution of healthy images is required to reliably identify new query images as within- or out-of-distribution relative to healthy samples, i.e. to identify them as anomalous or not. For image reconstruction, while GANs are most commonly used in an image-to-image fashion (e.g. for denoising image estimates while preserving realism), they also present the opportunity to learn the entire distribution of valid images, information which can be used as data-driven regularisation \citep{Reader2020}.

Given the potential uses for GAN models that approximate the entire distribution of tissue appearance, adapting them for different applications (e.g. tissue/modality combinations) is of great interest. See \cite{Quiros2021} for an example of the development of a high-quality GAN model in histology imaging. 


To summarise, to our knowledge this work is the first to train and evaluate GAN methods to model the appearance of healthy lung tissue in CT images in a fully 3D manner. Due to the extensive research required to produce and evaluate well-performing 3D GAN models that match the training data distribution both in appearance and in semantic features, we leave the evaluation of downstream tasks for future work. In the spirit of open-source biomedical research supporting code for this project is available online\footnote{\url{https://github.com/S-Ellis/healthy-lungCT-GANs}}.

The specific contributions of this work are:
\setlist{nolistsep}
\begin{enumerate}[noitemsep]
    \item we adapt GAN models from the literature to allow 3D output, and investigate their performance on the CT lung patch generation task with the commonly used \frechet\ inception distance (FID) metric and an observer study;
    \item we propose to use a minibatch discrimination technique (MDmin) and a method of increasing the effective batch size to improve performance over baseline methods; 
    \item we perform 3D, domain-specific analysis of the generated patches, and relate the 3D structure of output images to the latent space of GAN models; and
    \item these results provide baseline performance metric values for the lung CT patch generation task, serving as a reference for future work in this area.
\end{enumerate}

The structure of the rest of this paper is as follows: In Section 2, the principles of GANs are described, along with the details of the specific GAN methods that were employed in this study. Section 3 outlines the experimental set-up, including the data sampling, training process, and the evaluations that were performed. Sections 4 and 5 describe and discuss the results of the experiments and Section 6 summarises the conclusions of this work.

\section{Methods}

In this work we adapt \textblue{three GAN generator architectures to the 3D lung CT patch modelling problem: the widely investigated DCGAN \citep{Radford2016} representing a baseline approach, and styleGAN \citep{Karras2019} and bigGAN \citep{Brock2019} approaches which are more representative of the state-of-the-art in generator architectures. It is important to emphasise that these methods could not be used `out-of-the-box' and modification was required to allow 3D patch output. Therefore, we denote the methods investigated in this study as DCGAN3D, styleGAN3D, and bigGAN3D respectively.}

In general, a GAN model consists of a generator $G(\cdot)$ and a discriminator network $D(\cdot)$. The generator takes as input samples $z$ from the prior distribution $\mathbb{P}_z$ (usually a normal distribution), and outputs fake images $x_f$ so that $x_f = G(z)$ with $x_f \sim \mathbb{Q}$. The discriminator takes as input a minibatch of images $x$ from either (or both) the real and fake distributions, with real images denoted by $x_r$, and returns some per-sample value $D(x)$ which provides feedback on the `realism' of  $x$. The distribution of real images is denoted $\mathbb{P}$ and the aim of GAN training is for $\mathbb{Q} \rightarrow \mathbb{P}$ throughout the training process.

\subsection{\textblue{DCGAN3D generator}}

The deep convolutional GAN (DCGAN, \cite{Radford2016}) architecture was developed to be able to incorporate convolutions into a GAN model, since the original GAN of \cite{Goodfellow2014} used only fully-connected layers, which limited the size of the output images. The \textblue{DCGAN3D}\ generator used in this work is described in Appendix \ref*{sec:dcgan_archi}, adapted from the publicly available \textblue{2D}\ PyTorch implementation\footnote{\url{https://github.com/pytorch/examples/blob/master/dcgan/main.py}, accessed Feb 2020}. The total number of parameters was $\sim 24$ million. In summary, the latent code is expanded to a $4 \times 4 \times 4$ volume via convolution, and then this volume is progressively upsampled with transposed convolution layers until reaching the output size of $32 \times 64 \times 64$. Each transposed convolution operation is followed by a 3D batchnorm \citep{Ioffe2015} and a ReLU activation, except the last layer which has a tanh activation to bound the generated image in $[-1,1]$.

\subsection{\textblue{styleGAN3D generator}}

StyleGAN improves on other GAN techniques by using style-based image generation \citep{Karras2019}. The latent code is processed by a set of fully-connected layers and fed into a convolutional generator at multiple resolution scales to control the statistics of the activation maps using adaptive instance normalisation \citep{Huang2017}. This is a more powerful method of altering overall image appearance than learning across multiple convolutional-only layers.

The styleGAN generator also incorporates a mapping network which maps $z$ to a another, warped space, which has distribution $\mathbb{P}_w$. The aim is to allow the network to learn a more natural latent space, such that the $w$ vectors are better structured according to semantic image features \citep{Karras2019}.

\begin{figure*}[h]
\centering
\includegraphics[width=1\textwidth]{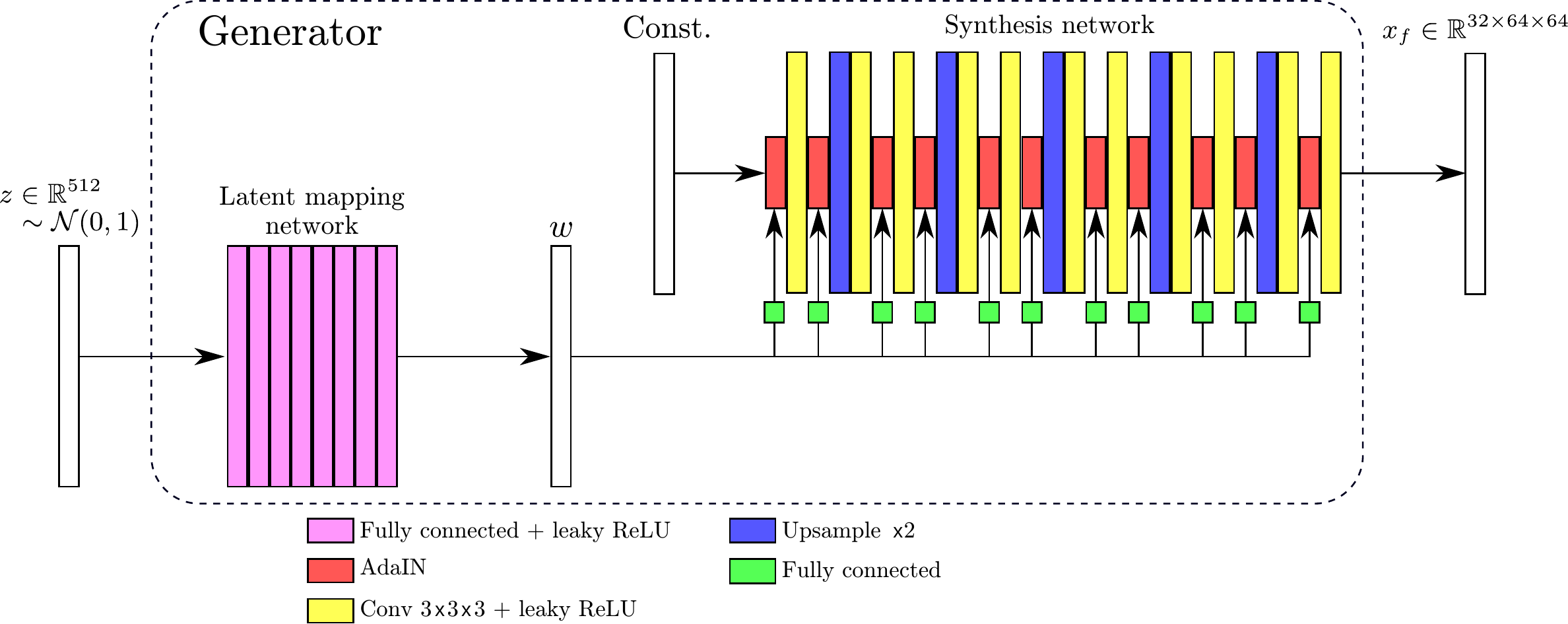}
\caption{Architecture of the styleGAN3D generator investigated in this work, based on the original implementation of \cite{Karras2019}. $z$ and $w$ are 512-element vectors, Const. is a constant image of size $1 \times 2 \times 2$, similar to that used in \cite{Karras2019}, and $x_f$ is a $32 \times 64 \times 64$ vox$^3$ image patch. AdaIN is adaptive instance normalisation, as described in \citep{Huang2017,Karras2018}.}
\label{fig:styleGAN_gen_archi}
\end{figure*}

Figure \ref{fig:styleGAN_gen_archi} shows schematically the architecture of the 3D styleGAN \textblue{(styleGAN3D)}\ patch generator implemented in this work, and Appendix \ref*{sec:stylegan_archi} describes the details of the latent mapping and synthesis networks. The total number of trainable parameters was $\sim 18$ million. \textblue{Note that in this work the noise injection channels of the original styleGAN architecture were omitted since we would wish to retain control of all relevant details of the generated images for the potential downstream tasks of anomaly detection and image reconstruction.}

\subsection{\textblue{bigGAN3D generator}}

\textblue{bigGAN is a competing high-quality GAN model first described by \cite{Brock2019}. By combining various architectural components that had previously been shown to aid GAN training, and systematically optimising the resulting architectures, bigGAN was demonstrated to allow stable training for high-resolution natural images. In particular, bigGAN inherits spectral normalisation \cite{Miyato2018} and self-attention layers \cite{Zhang2019} from previous GAN models. We adapted the publicly available bigGAN implementation\footnote{\url{https://github.com/ajbrock/BigGAN-PyTorch}, accessed Mar 2022} to 3D, with the full architecture listed in Appendix \ref{sec:biggan_archi}. Note that due to memory constraints the bigGAN3D architecture had only $\sim 4$ million trainable parameters.}

\subsection{Discriminator}

We used a simple convolutional feed-forward network as the discriminator backbone \textblue{for all methods}, as is common in both older and more recent GAN models \citep{Radford2016,Karras2018,Karras2019}. We also used minibatch discrimination (MD) to increase variability of GAN output. MD refers to a family of techniques which can be employed to avoid mode collapse (where many $x_f$ are similar) by allowing the discriminator to make comparisons between samples in a minibatch \citep{Salimans2016}. MD is particularly useful when the discriminator architecture is shallow, since shallow networks are intrinsically less able to detect mode collapse. 

\textblue{In this work we use a minimum-based MD method, denoted MDmin. Full details are given in Appendix \ref*{sec:mdmin}, but in summary the minimum perceptual difference between samples in a minibatch is provided to the discriminator to facilitate inclusion of whole-batch information. When mode-collapse has occurred for a generator model, this measure is small, allowing the discriminator to more easily detect that the minibatch is fake, providing a strong signal to the generator to diversify its samples.}

The details of the discriminator architecture, including the location of the MDmin layer are given in Table \ref{table:dis_archi}. The number of trainable parameters was $\sim 11.0$ million. Note that the MDmin layer is placed just before the final convolutional layer to allow the discriminator to quickly detect the MD signal in cases of mode collapse.

\textblue{We also note that the effectiveness of minibatch discrimination is limited by the maximum batch size that is computationally feasible. For this reason we use a heuristic trick to increase the effective batch size by pre-selecting the $k$ most similar (according to the MDmin) samples from a larger batch of $N$ samples, thereby allowing the discriminator to reliably see the most mode-collapsed samples for a particular model. We denote this training heuristic largeEBS (large effective batch size), and a full description is provided in Appendix \ref*{sec:mdmin}.}

\begin{table}[t]
\footnotesize
\begin{center}
\begin{tabular}{c}
\toprule
\textbf{Discriminator}\\
\midrule
Input: $x \in \mathbb{R}^{32 \times 64 \times 64} $\\
\midrule
Conv3d $2\times4\times4$, stride 2, pad $0 \times 1 \times 1$, no bias, 1$\rightarrow$64\\
\midrule
Leaky ReLU 0.2\\
\midrule
Conv3d $4\times4\times4$, stride 2, pad 1, no bias, 64$\rightarrow$128\\
\midrule
Leaky ReLU 0.2\\
\midrule
Conv3d $4\times4\times4$, stride 2, pad 1, no bias, 128$\rightarrow$256\\
\midrule
Leaky ReLU 0.2\\
\midrule
Conv3d $4\times4\times4$, stride 2, pad 1, no bias, 256$\rightarrow$512\\
\midrule
Leaky ReLU 0.2\\
\midrule
(Optional: MDmin) \\
\midrule
Conv3d $2\times4\times4$, stride 1, pad 0, no bias, 512/513$\rightarrow$1\\
\midrule
Output: $D(x) \in \mathbb{R}$\\
\bottomrule
\end{tabular}
\end{center}
\caption{Architecture of the 3D discriminator used in this investigation. Note that the MDmin layer (see text and Appendix \ref*{sec:mdmin} for description) was optional, and therefore the number of input channels to the subsequent layer is variable (512 without MDmin, 513 with).}
\label{table:dis_archi}
\end{table}

\subsection{Loss functions}

The original seminal work on GANs introduced by \cite{Goodfellow2014} proposed an intuitive approach where the discriminator aims to correctly classify real and fake images, while the generator aims to force the discriminator to misclassify fake images as real. Concretely, the losses for the two networks are:

\begin{align}
\mathcal{L}_{\textrm{D}} &= -\mathbb{E}_{x_r \sim \mathbb{P}} \left[ \log\left(
    \sigma(D(x_r))\right)\right] -\mathbb{E}_{x_f \sim \mathbb{Q}} \left[ \log\left(1 - 
    \sigma(D(x_f))\right)\right]
\label{eq:SGAN_loss_dis}
\end{align}
and 
\begin{equation}
\mathcal{L}_{\textrm{G}} = -\mathbb{E}_{x_f \sim \mathbb{Q}} \left[ \log\left(    \sigma(D(x_f))\right)\right]
\label{eq:SGAN_loss_gen}
\end{equation}
where $\sigma(\cdot)$ denotes the sigmoid function. Note that expectation values $\mathbb{E}[\cdot]$ are defined over the entirety of the distributions, but in practice these losses are calculatated batch-wise as an approximation. 

There are many alternative losses that have been proposed for training GANs, including Wasserstein distance \citep{Arjovsky2017,Gulrajani2017}, least-squares \citep{Mao2017}, Hinge loss \citep{Lim2017},  etc. Evaluating all such loss functions for a given application is a time-consuming exercise that we leave for future work. In this work we focus on one of the relativistic losses of \cite{Jolicoeur-Martineau2018}, which has been incorporated successfully into large-scale GAN models of medical images \citep{Quiros2021}. The equations for the relativistic losses are:

\begin{align}
\mathcal{L}_{\textrm{D}} &= -\mathbb{E}_{x_r \sim \mathbb{P}} \left[ \log\left(
    \sigma(\tilde{D}(x_r))\right)\right] -\mathbb{E}_{x_f \sim \mathbb{Q}} \left[ \log\left(1 - 
    \sigma(\tilde{D}(x_f))\right)\right]
\label{eq:RasGAN_loss_dis}
\end{align}

and 
\begin{align}
\mathcal{L}_{\textrm{G}} = -\mathbb{E}_{x_f \sim \mathbb{Q}} \left[ \log\left(
    \sigma(\tilde{D}(x_f))\right)\right] -\mathbb{E}_{x_r \sim \mathbb{P}} \left[ \log\left(1 - 
    \sigma(\tilde{D}(x_r))\right)\right]
\label{eq:RasGAN_loss_gen}
\end{align}

where:
\begin{align}
    &\tilde{D}(x_r) = D(x_r) - \mathbb{E}_{x_f \sim \mathbb{Q}} \left[D(x_f)\right] \nonumber \\
    &\tilde{D}(x_f) = D(x_f) - \mathbb{E}_{x_r \sim \mathbb{P}} \left[D(x_r)\right] 
\end{align}

Intuitively, these relativistic losses reframe the task of the discriminator and generator so that the discriminator now aims to assign higher output to real images than for fake ones. The generator then aims to `fool' the discriminator to do the opposite. Note that in this case the generator learns directly from both the fake and real images, whereas for the standard GAN losses (Equations \ref{eq:SGAN_loss_dis} and \ref{eq:SGAN_loss_gen}), the generator only receives direct feedback from the discriminator on the fake images.

\section{Experiments}

\subsection{Dataset}

We used the publicly available LUNA16 lung CT dataset \citep{Setio2017}, comprising 888 lung CT images, annotated by four radiologists. These annotations include malignancy scores and segmentations for nodules $\geq\SI{3}{mm}$ diameter. We chose to exclude nodules and pathological images from the present study to approximate the training of a GAN for a downstream anomaly detection task.

First, we exclude nodules by selectively rejecting sampled patches that contain any ground truth nodule voxels. Second, we considered all scans that had any single nodule with a median malignancy score (across radiologists) $\geq 4$ as diseased, excluding the whole CT volume from further use. Removing these malignant scans left 636 scans which were split into 509 training scans and 127 held out for future testing.

\subsection{Comparative Methods}

Using various combinations of the above GAN components results in a number of overall comparison methods for training our GAN. Table \ref{table:comparitiveMethods_noLoss} summarises the combinations we chose to compare in this work.


\begin{table}[t]
\footnotesize
\begin{center}
\begin{tabular}{l C{1.5cm} C{1.3cm} C{2.8cm} }
\toprule
Method abbreviation & Generator type & MDmin & Increased effective batch size\\
\midrule
\textblue{DCGAN3D-base} & DCGAN & \xmark & \xmark\\
\textblue{DCGAN3D-MDmin} & DCGAN & \cmark & \xmark\\
\textblue{DCGAN3D-MDmin-largeEBS} & DCGAN & \cmark & \cmark\\
\textblue{styleGAN3D-base} & styleGAN & \xmark & \xmark\\
\textblue{styleGAN3D-MDmin} & styleGAN & \cmark & \xmark\\
\textblue{styleGAN3D-MDmin-largeEBS} & styleGAN & \cmark & \cmark\\
\textblue{bigGAN3D-base} & \textblue{bigGAN} & \textblue{\xmark} & \textblue{\xmark}\\
\textblue{bigGAN3D-MDmin} & \textblue{bigGAN} & \textblue{\cmark} &\textblue{\xmark}\\
\textblue{bigGAN3D-MDmin-largeEBS} & \textblue{bigGAN} & \textblue{\cmark} & \textblue{\cmark}\\
\bottomrule
\end{tabular}
\end{center}
\caption{Summary of GAN methods compared in this study, and their constituent parts.}
\label{table:comparitiveMethods_noLoss}
\end{table}

\subsection{Model Training}

\begin{figure*}[h]
\centering
\includegraphics[width=0.8\textwidth]{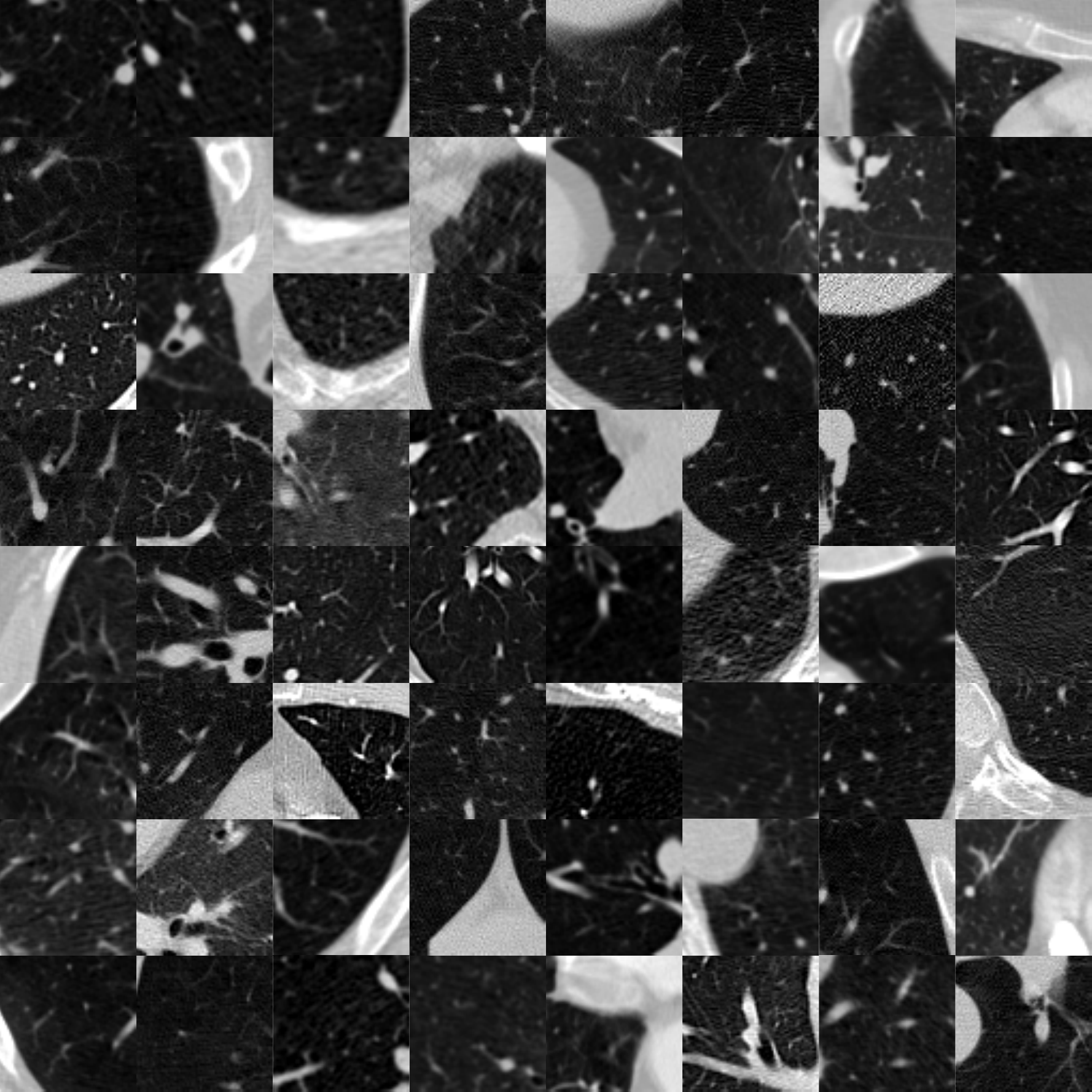}
\caption{Example real samples from the training dataset (central slices only). Note the high variability in terms of features such as lung wall placement and vessel structure.}
\label{fig:real_samples}
\end{figure*}

The training of GAN model involves sampling from the real image distribution, as well as generating fake images using the generator. As shown above, the GAN output was of size $32 \times 64 \times 64$ vox$^3$. Therefore patches sampled from the real CT images were also of this size. Patches were sampled from within the lung using the LUNA16 provided lung masks, excluding all voxels annotated as nodules as described above. Figure \ref{fig:real_samples} show a sample of real patches (centre slice only), highlighting the high variability in lung wall position and vessel structure. Note that resampling to a standard voxel size and other data augmentations (translation, rotation, etc) were not performed.

Unless stated otherwise, for a given CT image, 672 patches were sampled at a time, windowed to $[-1000,400]$ HU, and then rescaled to $[-1,1]$. These patches were then used to create minibatches of size 48, with 14 minibatches per image. For each update of the discriminator, this batch of 48 real images is passed through to obtain $D(x_r)$, followed by a batch of 48 fake images to get $D(x_f)$. A similar sampling process is followed for the generator updates. A single pass through all 509 training CT scans (i.e. 7126 iterations) is considered an epoch of training. For each epoch, patches are sampled separately, ensuring that with high probability no single patch is repeated during the training.

For all GAN models, the generator and discriminator learning rates were set the same at $0.0001$, using the Adam optimiser \citep{Kingma2015} with $\beta_1=0.5$. The discriminator and generator were each updated once per iteration. Training was run for 20 epochs using either an NVIDIA GeForce GTX 1080 Ti or an NVIDIA TITAN Xp GPU. 

Finally, for the \textblue{styleGAN3D}\ approach, so-called style-mixing regularisation was used \citep{Karras2019} which swaps two latent codes $w_1$ and $w_2$ at a random depth in the convolutional generator.  Since the GAN training process still encourages the output to be realistic, the effect is to decouple the effect of the latent code at different scales (see \cite{Karras2019} for further details). \textblue{Note that in comparison to the standard 2D styleGAN method of \cite{Karras2019}, progressive growing of images was not found to be necessary for styleGAN3D due to the small relative size of the output patches compared to the convolutional kernel sizes.}

\subsection{Automatic Model Evaluation with FID}

GAN models were evaluated throughout training with the commonly-used \frechet\ inception distance (FID) which was proposed and explained in detail by \cite{Heusel2017}. In brief, the FID aims to measure the similarity between the distributions of real training images ($\mathbb{P}$) and fake, GAN produced images ($\mathbb{Q}$). To do this, a large number ($\sim 5000$--$10000$) of images from each distribution is sampled. These images are passed through a pre-trained network to obtain the activations at a deep layer, which can be assumed to be Gaussian in distribution. Measuring the \frechet\ distance \cite{} between these two distributions provides a measure of similarity between the original distributions. The FID provides a measure of both average image quality (i.e. `realism') and variability, and so it is sensitive to mode collapse.

As mentioned, FID uses a pre-trained network to provide higher-level representations of the input images. In common practice, this network is \textblue{an Inception network pre-trained on a natural image task \citep{Heusel2017}, which presents two considerations for use in the current investigation.} Firstly, the pre-trained network accepts only 2D input. Therefore, when calculating FID for our 3D GANs, we use only the central slices of generated images. Secondly, since the data domain is different (natural images vs lung CT patches), the \textblue{transferability of the standard Inception network to lung CT could be questionable. However, we note that calculating the FID between two random sets of real images produces a low FID value of $\leq 4$, suggesting that producing a low FID score is a valid target for GAN models of lung CT patches.} 

We used FID to measure GAN performance throughout training, with five training runs performed for each method listed in Table \ref{table:comparitiveMethods_noLoss}. For each method we selected the lowest-FID model from across the five training runs as the best. These lowest FIDs were then compared between architectures, and the corresponding models were retained for further investigation and characterisation.

\subsection{Observer Study}

To complement the results of the automated model evaluation, we also performed a human observer study on the best performing 3D GAN models. One human observer (SE) was provided 200 2D image slices per model, split 50:50 into real and fake. The task was to classify real and fake images, and this was repeated three times to assess intra-observer variability. Receiver operating characteristic (ROC) analysis of the results was then performed.

\subsection{3D Domain-Specific Analysis}

\textblue{The analyses and model selection described above were performed only in 2D due to the reliance of the standard FID on a 2D Inception network}, and the infeasibility of performing large observer studies with 3D patches. To investigate the 3D, domain-specific structure of the generated patches, additional analysis was performed.

\textblue{Firstly, although the conventional FID measurement uses a trained 2D Inception network, there has been research proposing the use of a 3D network to enable a 3D FID score \citep{Sun2020}. In this case the network from which activations are retrieved is a 3D ResNet pre-trained on different medical imaging segmentation datasets \citep{Chen2019}. Transfer learning experiments showed that the learnt features in such a network were beneficial for a range of tasks, including segmentation of lung tissue in CT images \citep{Chen2019}. We use one model provided by \cite{Chen2019} to calculate a 3D FID (FID3D) for our lung tissue GANs. Specifically, we use a pre-trained ResNet-10 model, and obtain the output activation maps, which have dimensions of $512\times4\times8\times8$ when given $1\times32\times64\times64$ input. To avoid excessive memory usage we spatially average these maps to obtain a $512$ element feature per input patch. Processing $\sim 10000$ real and fake images allows the \frechet\ distance to be measured between real and fake datasets as for standard 2D FID.}

\begin{figure*}
\centering
\includegraphics[width=0.8\textwidth]{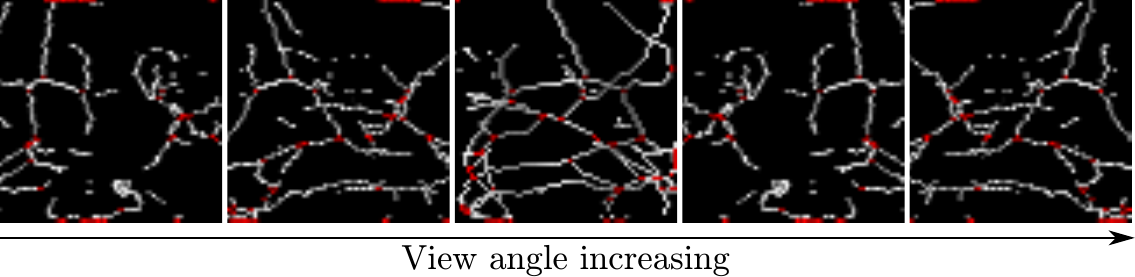}
\caption{Example 3D skeleton MIPs and branch points extracted from a single image patch. Multiple viewing angles are displayed for clarity, and detected branch points are highlighted in red.}
\label{fig:example_branch_points}
\end{figure*}

\textblue{Secondly, we note that if the distributions of real and fake images are close, then we expect that the distributions of any quantity derived from those images should also be close, i.e. comparing the distributions of any characteristic of real and fake images is valid. Beyond this, using a quantity relevant for the task at hand is preferable. To the best of our knowledge, little research has been performed on the expected appearance of healthy lung tissue in CT image patches. Therefore we opted to analyse the appearance of a prominent feature of healthy lung tissue patches: the vasculature. Specifically we chose to model the complexity of the vasculature by finding the number of branch points in an automated fashion. 3D patches were skeletonised using standard python scikit-image library functions in order to provide a sparse representation of the 3D structure. These 3D skeletons were visually assessed using 3D rotating maximum intensity projections (MIPs). The number of branch points was automatically detected, and the distribution of branch points across $\sim$10k images was recorded. Figure \ref{fig:example_branch_points} shows an example skeletonisation of a generated patch, with automatically detected branch-points highlighted.}

Finally, the relationship between the latent space and the distribution of branch points was investigated using the UMAP dimensionality reduction technique \citep{McInnes2020}. Using the freely available implementation of UMAP\footnote{\url{https://umap-learn.readthedocs.io/en/latest/index.html}, accessed Oct 2021}, the 512-dimensional latent spaces for each method ($z$ for DCGAN3D and \textblue{bigGAN3D}, $w$ for styleGAN3D) were transformed into a two-dimensional embedded space. The embedded space was calculated using 50000 random samples to mediate the curse of dimensionality. For 1000 of these samples, the original latent vectors were passed through the corresponding GAN model to produce fake images which could then be analysed as above to find the number of branch points. Latent space structure could then be visualised by labelling each point in the embedded latent space with its corresponding number of branch points.




\section{Results}

\subsection{Model Selection}

 \textblue{Table \ref{table:minFIDs} shows the minimum observed FIDs for each model, as well as the average and standard deviation (SD) of the minimum over multiple runs. The SD of the minimum FID over the five runs provides a measure of training stability, since it reflects the ability of the method to provide similar results under different initialisations and sequences of training images. The significance of the effects of the MDmin and larger effective batch size techniques was established by performing Welch's $t$-test on the minimum FID in the five training runs for pairs of methods using the same generator architecture (Figure \ref{fig:FID_sig}). FID values and network losses throughout the training are provided in Appendix \ref*{sec:suppl_results}.}

\textblue{For the DCGAN3D method, including the MDmin layer provides a small, non-significant ($p=0.64$) improvement, reducing the minimum FID from $89.0$ to $80.3$ and the average minimum FID only from $93.1\pm3.1$ to $91.0\pm9.0$. The increased SD over five runs suggests that the modest performance improvement comes at the cost of reduced training stability. In addition, we found that the performance of DCGAN3D-base was not significantly affected by batch size (Appendix \ref{sec:suppl_results}).

On the other hand, MDmin was observed to have a significant effect for the styleGAN3D architecture. Without the MDmin layer, the minimum FID observed was $122.3$, with an average minimum of $136.9\pm10.7$. Including the MDmin layer reduced this to a minimum of $41.0$ with an average of $44.8\pm4.2$ ($p<<0.05$); a reduction in both mean and SD representing improved training stability. For the bigGAN3D architecture, the use of MDmin reduced FID from $93.8\pm5.4$ to $58.4\pm5.8$ which was a significant improvement ($p<<0.05$).}

\textblue{The heuristic method to increase the effective batch size (largeEBS) provided non-significant improvements for the DCGAN3D and bigGAN3D architectures, although for the styleGAN3D architecture, the minimum FID reduced from $41.0$ with just the MDmin layer to $35.3$ when using the increased effective batch size, reflecting a significant reduction in mean minimum FID ($p<0.05$). }


\begin{table}[t]
\footnotesize
\begin{center}
\begin{tabular}{l C{1.5cm} C{2.8cm} }
\toprule
 & Minimum FID & Average minimum FID over five runs\\
\midrule
\textblue{DCGAN3D}-base & 89.0 & 93.1$\pm$3.1 \\
\textblue{DCGAN3D}-MDmin & 80.3 & 91.0$\pm$9.0 \\
\textblue{DCGAN3D}-MDmin-largeEBS & 80.5 & 86.8$\pm$5.1 \\
\textblue{styleGAN3D}-base & 122.3 & 136.9$\pm$10.7 \\
\textblue{styleGAN3D}-MDmin & 41.0 & 44.8$\pm$4.2 \\
\textblue{styleGAN3D}-MDmin-largeEBS & \textbf{35.3} & \textbf{38.5$\pm$4.3} \\
\textblue{bigGAN-base} & \textblue{86.5} & \textblue{93.8$\pm$5.4} \\
\textblue{bigGAN-MDmin} & \textblue{52.4} & \textblue{58.4$\pm$5.8} \\
\textblue{bigGAN-MDmin-largeEBS} & \textblue{51.6} & \textblue{57.4$\pm$6.3}\\
\bottomrule
\end{tabular}
\end{center}
\caption{Minimum FIDs for each GAN model listed in Table \ref{table:comparitiveMethods_noLoss}. Both the minimum across five runs, and the mean$\pm$SD are provided. }
\label{table:minFIDs}
\end{table}

\begin{figure*}[h]
\centering
\includegraphics[width=1\textwidth]{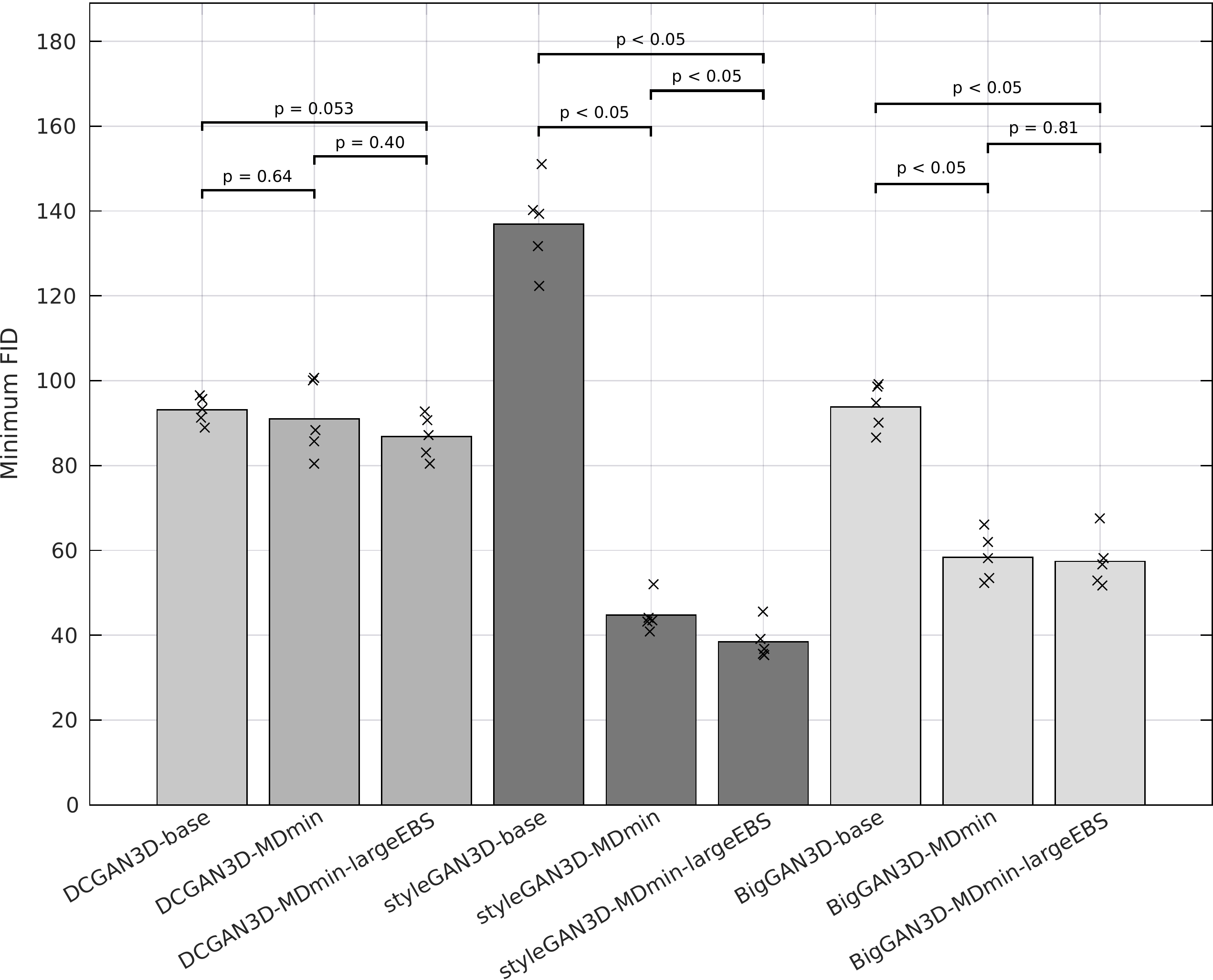}
\caption{The minimum FID across five runs for each method, along with $p$-values measured for pairwise comparisons within each architecture using the Welch's $t$-test.}
\label{fig:FID_sig}
\end{figure*}

\begin{figure}
\centering
\includegraphics[width=1\textwidth]{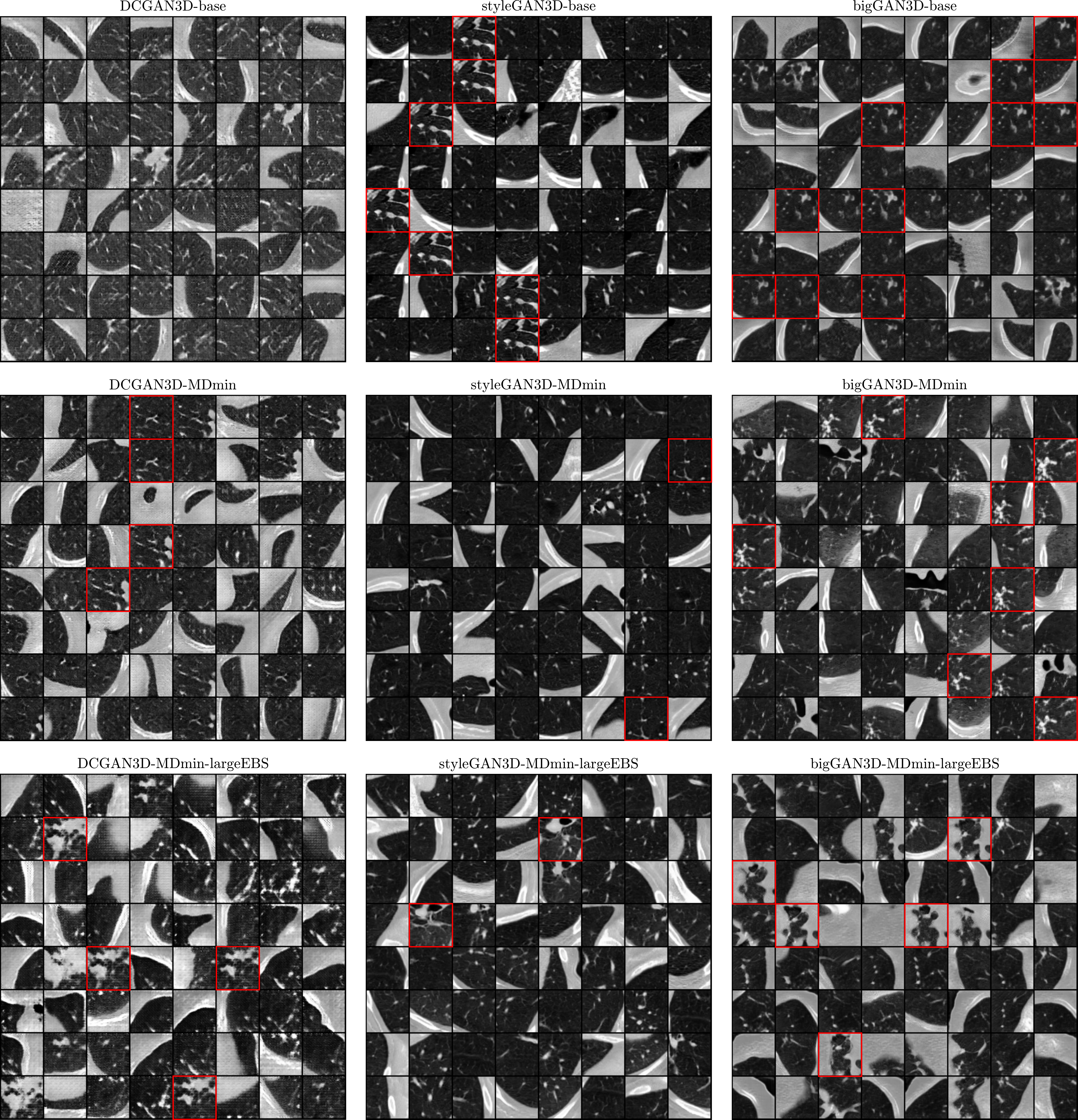}
\caption{Uncurated samples from each of the GAN methods in Table \ref{table:comparitiveMethods_noLoss}, corresponding to the minimum FID models in Table \ref{table:minFIDs}. Example mode-collapsed images are highlighted with red squares. Note that the images shown are the central slices from the 3D GAN output.}
\label{fig:minFID_samples}
\end{figure}

\textblue{Figure \ref{fig:minFID_samples} shows some samples for the minimum FID models for each of the GAN methods. The DCGAN3D-base method shows a variety of images, but the image quality is poor, with periodic artefacts visible. On the other hand, the styleGAN3D-base and bigGAN3D-base methods produce high-quality images, but with severe mode collapse. The DCGAN3D-MDmin and DCGAN3D-MDmin-largeEBS methods improve on image quality compared to DCGAN3D-base, but some mode collapse has arisen. This seems to contradict the purpose of the MDmin and largeEBS techniques, though it should be noted that the minimum FID for the best DCGAN3D-base training run occurred early in training, whereas the DCGAN3D-MDmin method reached a minimum at a later point. Because of this, a direct comparison of the variability of the two methods is not appropriate, since mode collapse usually occurs later in training.}

\textblue{In contrast to DCGAN3D, the styleGAN3D-MDmin model reduces mode collapse while retaining image quality compared to the styleGAN-base method. For bigGAN3D, the MDmin did not visibly reduce mode collapse to the same extent as for styleGAN, despite the significant drop in FID recorded in Table \ref{table:minFIDs}. However, image quality seems to be qualitatively improved, with more structure visible within the lung parenchyma. Similarly to the DCGAN3D methods, this is likely due to the early occurrence of the minimum FID for the bigGAN3D-base method, before image quality was stable. Using the largeEBS trick seems to further reduce mode collapse for the styleGAN architecture, whereas for the bigGAN3D architecture the effect was negligible.}

\begin{figure}[h]
\centering
\includegraphics[width=1\textwidth]{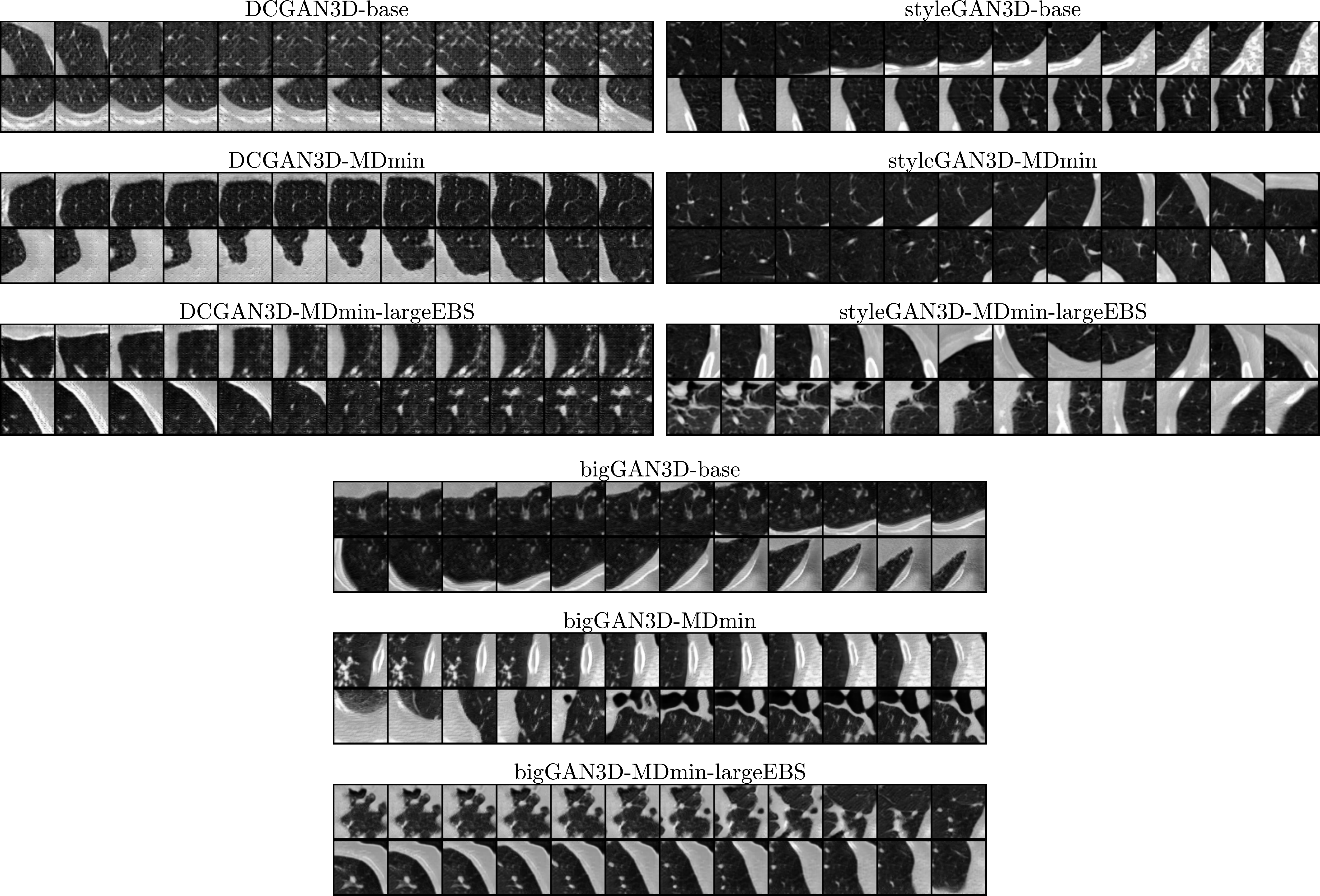}
\caption{Latent space interpolation for each minimium FID model. In each case two independent interpolations are shown. \textblue{Note that the DCGAN and bigGAN interpolations consisted of a spherical interpolation in the $z$-space, whereas for styleGAN methods linear interpolation was performed in $w$-space.}}
\label{fig:minFID_interps}
\end{figure}

\textblue{Finally, Figure \ref{fig:minFID_interps} shows interpolations between two randomly sampled latent codes for each of the minimum FID models. Note that DCGAN3D and bigGAN3D methods were spherically interpolated in $z$-space, whereas styleGAN methods were linearly interpolated in $w$-space, as described in the original description of the method \citep{Karras2019}. A smooth latent space is a desirable quality of GAN models, which should be reflected by smooth transitions between sampled images. The DCGAN3D methods provide smooth transitions despite their poor image quality. The bigGAN3D methods also exhibit smoothness, with a superior image quality. In general the styleGAN3D interpolations in $w$-space are smooth, although some discontinuities/non-smooth regions are present, possibly due to interpolations traversing low-density regions of $w$-space. }

\subsection{Observer Study}

\begin{figure}
\centering
\includegraphics[width=1\textwidth]{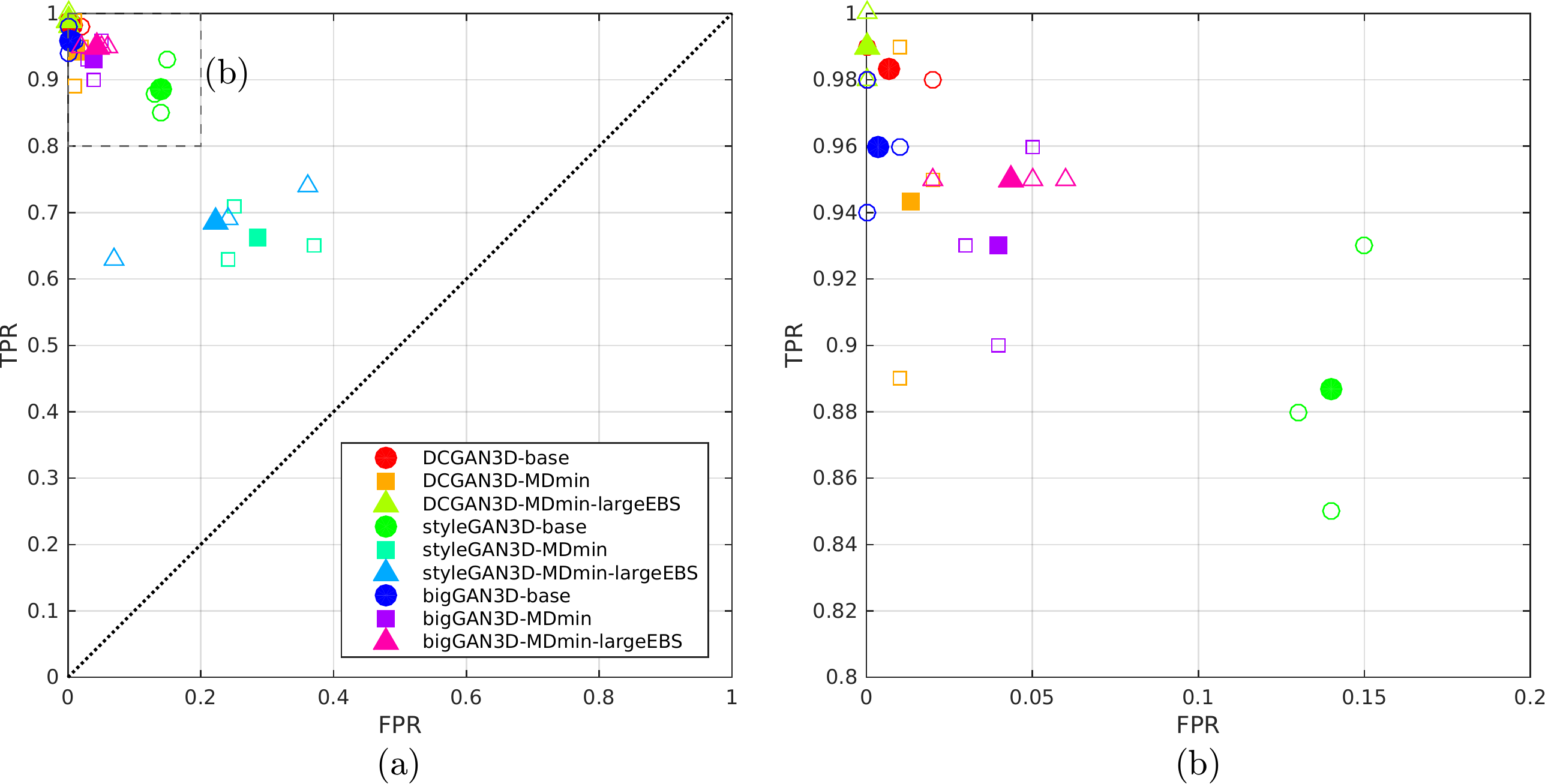}
\caption{\textblue{Observer study ROC graph. Since good generative models should produce samples indistinguishable from real samples, GAN performance increases closer to the TPR$=$FPR line, and worsens towards TPR$=1$, FPR=$0$ (a) The average performances over three experiments are represented by solid markers for each method, with individual experiments shown with unfilled markers. (b) A zoomed version is also provided to more clearly show the low-FPR, high-TPR area of ROC space.}}
\label{fig:roc}
\end{figure}

\textblue{Figure \ref{fig:roc} shows the results of the observer study. The DCGAN3D methods, suffering from periodic artefacts and poor image quality, were easily distinguishable from the real images, with $\textrm{FPR} < 0.025$ and $\textrm{TPR}>0.88$ for all models. Slightly improved performance was observed using the bigGAN3D models, with $\textrm{FPR}=0.04$ and $\textrm{TPR}=0.93$ for the bigGAN3D-MDmin method, but some remaining convolution artefacts made the fake images conspicious compared to real samples. The styleGAN3D-base method performed better than all the DCGAN3D and bigGAN methods, but the mean observer performance was still high at $\textrm{FPR}=0.14$ and $\textrm{TPR}=0.89$. When including the MDmin layer, observer performance dropped to $\textrm{FPR}=0.29$ and $\textrm{TPR}=0.66$, representing improved model performance. The large effective batch size heuristic worsened styleGAN3D performance slightly, yielding $\textrm{FPR}=0.22$ $\textrm{TPR}=0.69$.}

\subsection{3D Analysis}

\textblue{The FID3D score for each of the models is shown in Table \ref{table:FID3D}. Similarly to the conventional 2D FID scores in Table \ref{table:minFIDs}, the best performing method overall is the styleGAN3D-MDmin-largeEBS model which produced a value of 0.57 for the FID3D metric. Interestingly, the effect of the largeEBS method is different in 2D FID compared to 3D FID. For example, the largeEBS reduces 2D FID for the bigGAN3D, compared to using MDmin alone, whereas in terms of FID3D, it rises considerably. We hypothesise that this is because the bigGAN3D architecture has fewer trainable parameters, which means that the MDmin method alone may be fully utilising network capacity. Pushing further with the largeEBS method may cause the training to become more unstable, resulting in worse 3D structure. We leave a full comparison of 2D and 3D FID metrics for future work.}

\begin{table}[t]
\footnotesize
\begin{center}
\begin{tabular}{l C{1.5cm}}
\toprule
 & \textblue{FID3D} \\
\midrule
\textblue{DCGAN3D-base} & \textblue{4.40}\\
\textblue{DCGAN3D-MDmin} & \textblue{4.40} \\
\textblue{DCGAN3D-MDmin-largeEBS} & \textblue{1.30} \\
\textblue{styleGAN3D-base} & \textblue{3.97} \\
\textblue{styleGAN3D-MDmin} & \textblue{2.90} \\
\textblue{styleGAN3D-MDmin-largeEBS} & \textblue{\textbf{0.57}}  \\
\textblue{bigGAN3D-base} & \textblue{4.24}  \\
\textblue{bigGAN3D-MDmin} & \textblue{1.88}  \\
\textblue{bigGAN3D-MDmin-largeEBS} & \textblue{6.05}\\
\bottomrule
\end{tabular}
\end{center}
\caption{\textblue{The FID3D scores for each GAN model investigated. Note that these scores correspond to the models used elsewhere, i.e. the lowest FID obtained throughout training over five independent runs.}}
\label{table:FID3D}
\end{table}

\begin{figure}[h]
\centering
\includegraphics[width=1\textwidth]{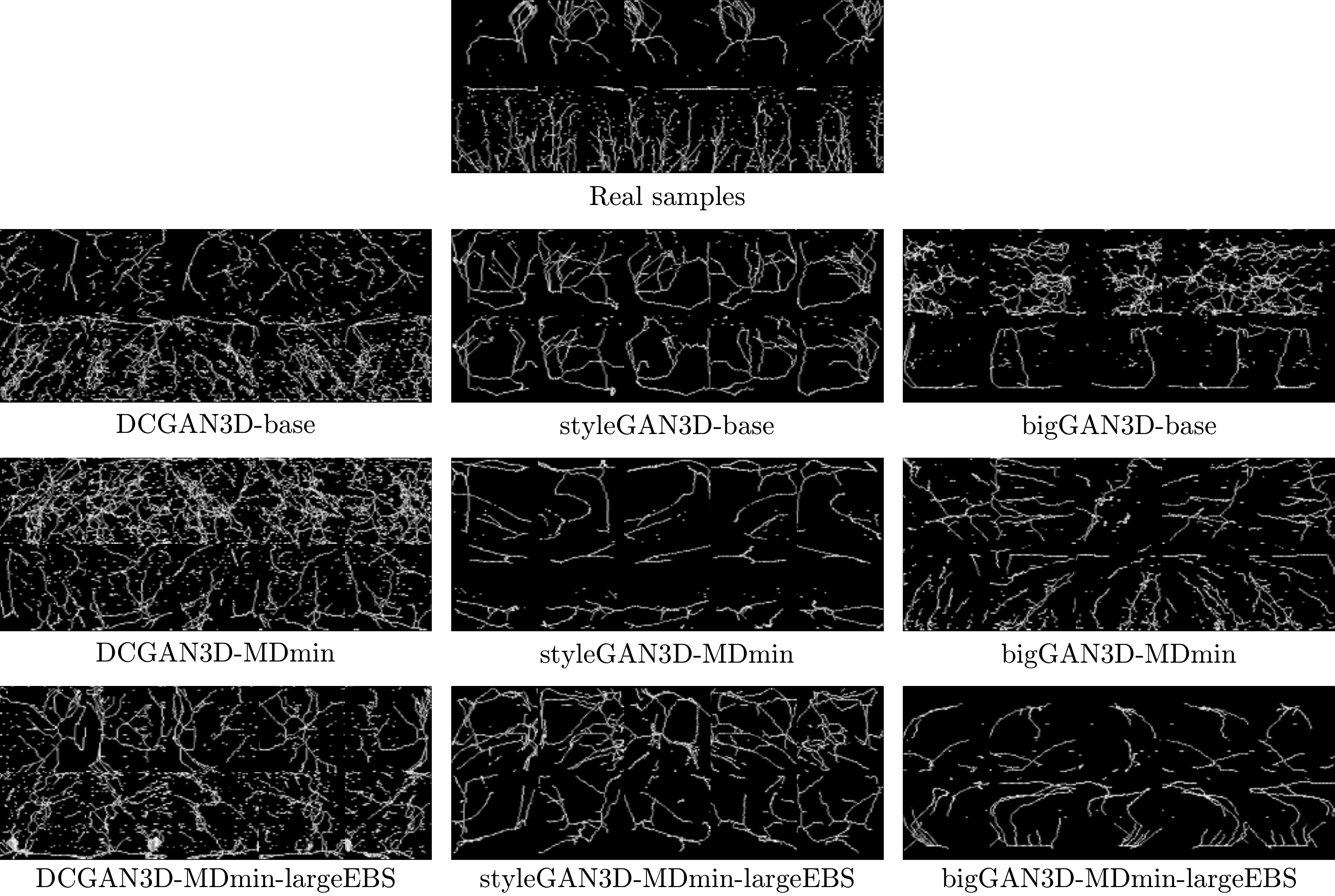}
\caption{Visualisation of 3D structure of generated samples with 3D skeletons. Each method shows two separate 3D skeletons at 10 view angles. The \textblue{DCGAN3D}\ methods produce noisy skeletons, whereas the \textblue{styleGAN3D}\ methods produce cleaner skeletons with a clearer structure.}
\label{fig:skel3d}
\end{figure}

Figure \ref{fig:skel3d} shows some example skeletons calculated from random outputs of each GAN model. The \textblue{DCGAN3D}\ models provide noisy skeleton images, due to the poor quality of the generated samples. In contrast the \textblue{styleGAN3D and bigGAN3D}\ outputs produce cleaner skeletons with clearer structure, more closely resembling real images. 

\begin{figure}[h]
\centering
\includegraphics[width=0.7\textwidth]{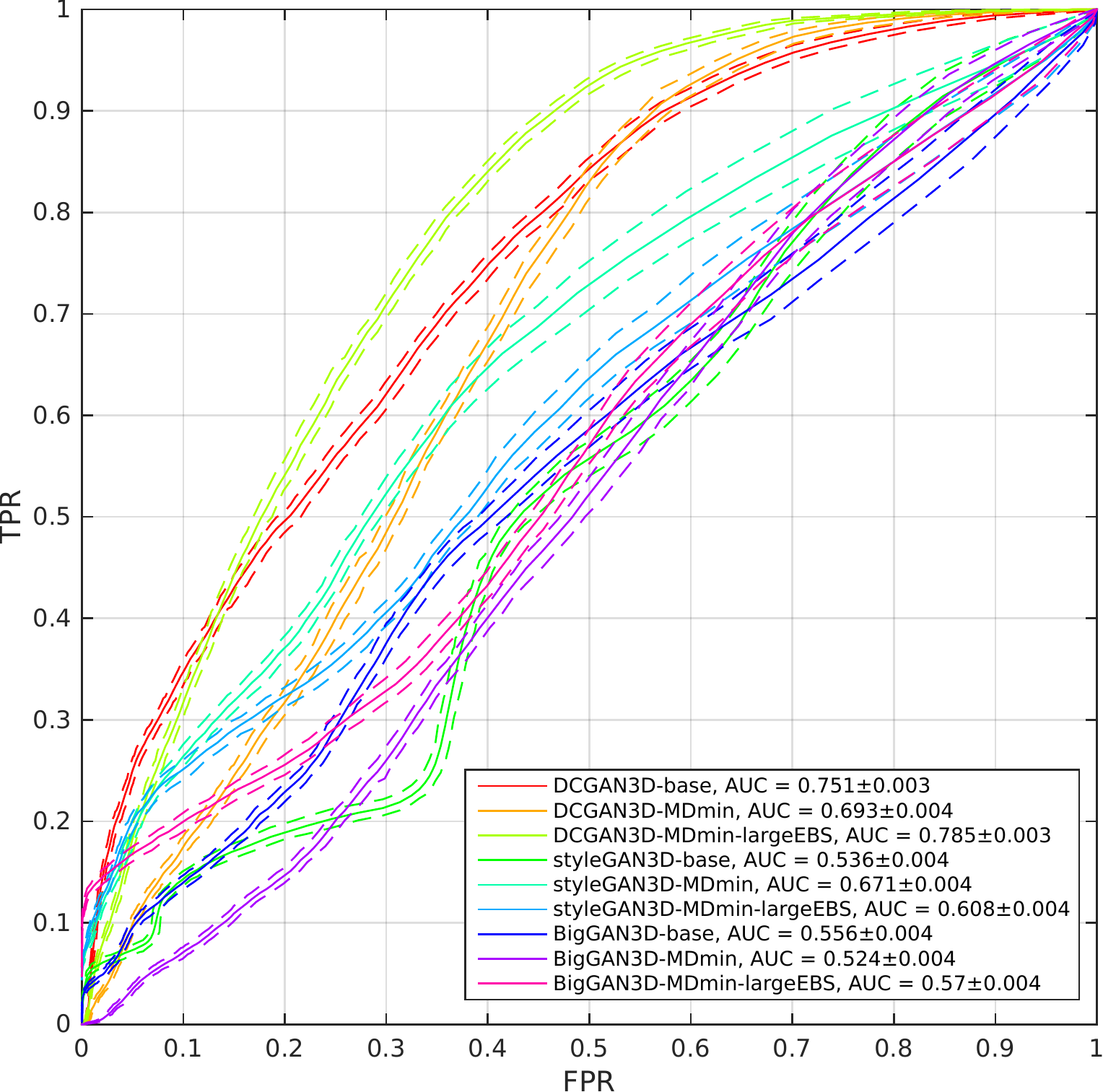}
\caption{ROC curves calculated with the distribution of 3D branch points in real and fake images, for each of the GAN methods. Dashed lines show 95\% percentile bootstrap confidence intervals, and AUCs are displayed $\pm$ the standard deviation of bootstrapped AUCs. \textblue{The bigGAN methods provide better distributions of branch points than the DCGAN or styleGAN methods, reflected by AUC values closer to 0.5.}}
\label{fig:branchROC}
\end{figure}

\textblue{To analyse the distributions of 3D branch points in real and fake images, ROC curves were produced for each method by varying a threshold number of branch points and recording the true and false positive rates (TPR and FPR) of the resulting binary classification. Methods that generate fake images with a similar distribution of branch points to the real data will produce areas under the ROC curve (AUCs) closer to 0.5, and vice versa.}\ Figure \ref{fig:branchROC} shows the ROC curves each of the methods. The DCGAN methods provide AUCs of $0.750\pm0.003$ (DCGAN-base), $0.693\pm0.004$ (\textblue{DCGAN3D}-MDmin), and $0.784\pm0.003$ (\textblue{DCGAN3D}-MDmin-largeEBS), demonstrating that the DCGAN architecture is unable to produce a realistic distribution of branch points. In contrast, the \textblue{styleGAN3D}\ architecture produces distributions of branch points closer to the real one, reflected by AUC values of $0.535\pm0.004$ for \textblue{styleGAN3D}-base, $0.671\pm0.004$ for \textblue{styleGAN3D}-MDmin, and $0.608\pm0.004$ for \textblue{styleGAN3D}-MDmin-largeEBS. \textblue{Overall, the bigGAN3D methods provided the best performance, with a lowest AUC of $0.524\pm0.004$.}

\begin{figure}[h]
\centering
\includegraphics[width=0.95\textwidth]{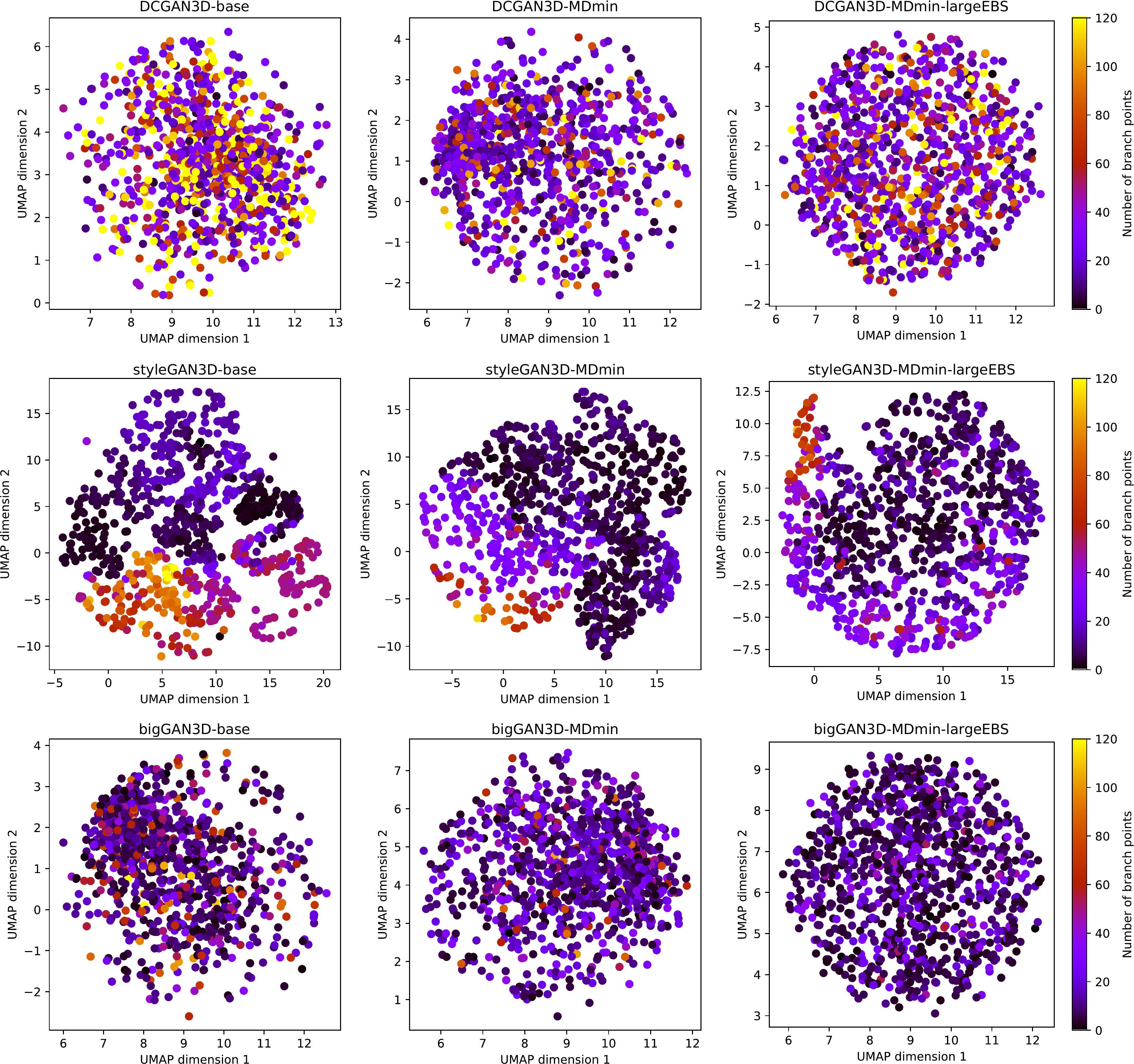}
\caption{\textblue{Visualisation of latent space using UMAP. Colour corresponds to the number of branch points measured by the previously described pipeline. For the DCGAN3D and bigGAN3D architectures the position of a sample in UMAP space does not relate to the number of branch points in the corresponding generated image. On the other hand, for the styleGAN3D methods images with similar numbers of 3D branch points are clustered together. This clustering is particularly strong for images containing a large number of branch points (>~70). This suggests that the styleGAN3D models were able to learn latent space representations that relate to semantically meaningful quantities. Note that the colour scale is saturated at 120 for clarity.}}
\label{fig:UMAP_branches}
\end{figure}

\textblue{The UMAP visualisation of the latent spaces, and their relationship to the number of branch points are shown in Figure \ref{fig:UMAP_branches}}. For the \textblue{DCGAN3D and bigGAN3D}\ methods, the UMAP-embedded space shows no structure since the $z$-space is by design distributed normally. Furthermore, no correspondence is apparent between the position of a point in the embedded space and the number of branch points in the corresponding generated image. Conversely, the \textblue{styleGAN3D}\ methods show both structure in the shape of the $w$-space, and a correspondence between position and number of branch points. \textblue{This suggests that higher level concepts are contributing to the structure of the full latent space, agreeing with previous literature on the use of styleGAN for medical image generation \citep{Quiros2021}. This is an interesting result, opening the possibility for further investigation into the structure of the latent space with respect to other semantically meaningful features in future work. 

It is also important to emphasise that the lack of correspondence between UMAP embeddings and the number of branch points for the bigGAN3D and DCGAN3D methods does not suggest that the full $512$-dimensional latent spaces for these methods are not ordered with respect to the number of branch points, just that any such correspondence is not preserved under the UMAP transformation. Additional analysis to explore whether such correspondence exists for the DCGAN3D and bigGAN3D methods remains for future work.}

\section{Discussion}

\textblue{The aim of this work was to investigate the applicability and transferability of existing 2D GAN models for the problem of generating 3D patches to model the appearance of healthy lung tissue in pulmonary CT images. Although there has been much research using GANs to generate images of pulmonary nodules \citep{Liu2018,Jin2018,Xu2019,Yang2019,Bu2021,Suh2021,Wang2021}, we believe we are the first to take the next step to specifically model healthy lung tissue. 

Three architectures were selected from the literature: DCGAN, styleGAN and bigGAN. Results show that the three methods have very different characteristic behaviour. From a visual inspection of generated images, the DCGAN3D-base model was capable of generating a higher variety of images than the styleGAN3D and bigGAN networks (Figure \ref{fig:minFID_samples}), but at the cost of reduced image quality reflected by poor performance in a human observer study (Figure \ref{fig:roc}) and a higher FID (Table \ref{table:minFIDs} and Figure \ref{fig:FID_sig}). In contrast, the styleGAN3D-base architecture produced more realistic images that performed better in the observer study and had a more realistic 3D structure (Figures \ref{fig:skel3d} and \ref{fig:branchROC}), but at the cost of severe mode collapse that required a minibatch discrimination (MDmin) layer to mitigate. In terms of the FID metric the styleGAN3D methods with MDmin performed best even taking into account remaining mode collapse (Table \ref{table:minFIDs} and Figure \ref{fig:FID_sig}).  Use of the latent mapping network was crucial to the performance of the styleGAN3D methods: omitting it caused training to become unstable and greatly increased the minimum FID (Appendix \ref{sec:suppl_results}). 

Despite the strong performance of the styleGAN3D methods in terms of FID and a human observer study, the bigGAN methods were best at learning the distribution of a semantically meaningful quantity (Figure \ref{fig:branchROC}), even with a smaller number of trainable parameters ($\sim4$ million for bigGAN3D vs $\sim 18$ million for styleGAN3D). These results suggest a number of possible improvements to be made in future work: 1) training a bigGAN3D with the same number of parameters as styleGAN3D may lead to better performance in the other metrics while maintaining the distribution of semantically important features, and 2) combining components from styleGAN3D and bigGAN3D architectures, for example by upgrading the bigGAN architecture to allow style-based generation, could yield a method superior in all metrics.}

\textblue{Beyond the networks and architectural components investigated in this work, many more possibilities remain, including the exploration of upgraded styleGAN approaches \citep{Karras2020,Karras2021} and autoencoder models such as the high-quality VQ-VAE family \citep{Oord2017,Razavi2019}. The methodology and results presented here represent an important baseline for such future investigations.}

\textblue{Another area for future research is the inclusion of positional information in the GAN models. Currently, our models treat the generation task in a `bag-of-patches' fashion, i.e. ignoring the potentially important spatial context for each patch. By including positional information such as in \cite{Lin2020,Dosovitskiy2021}, the models could be trained to learn the distribution of healthy tissue appearance at each position within the lung, utilising the prior information to help training. However, these methods would need a greater amount of training data, and would take longer to train.}

\textblue{In any case the use of patch-based generators as investigated in this work may hinder application of such methods to a downstream task. Although the query image can be decomposed into patches and then passed through the anomaly detection/reconstruction pipeline, ensuring continuity/consistency between the overlapping output patches may be difficult. In theory a GAN that has learnt the full data distribution should be shift-invariant (given that translated patches remain valid patches), but sub-optimal GAN models are likely to suffer artefacts that would need addressing in future work.}

\textblue{A potential solution to this issue would be to perform whole-volume image generation, completely obviating the need to work with patches. While there has been recent research into modelling large 3D volumes with GANs \citep{Uzunova2020,Sun2020,Pesaranghader2021,Hong2021}, the modelling of entire 3D volumes will require a much larger amount of data in order to fully model the data distribution. The use of smaller datasets has been shown to yield whole-lung GAN models suitable for a data augmentation task \citep{Pesaranghader2021}, but there is currently insufficient evidence that the models capture well the data distribution everywhere, as would be required for an anomaly detection or image reconstruction task. Nonetheless, these models can be evaluated under the pipeline introduced in this work by firstly generating a number of whole-volume images, and then breaking these into patches for subsequent analysis. }

%



\section{Conclusions}

GAN models are of interest in anomaly detection and image reconstruction tasks in a variety of medical imaging applications. However, due to variations in the underlying structure of datasets, GAN models must be re-adapted, re-trained, and re-assessed for each new application, which is a non-trivial task. 

\textblue{In this work we adapted three GAN models from the natural image literature to the task of 3D healthy lung CT patch generation, and evaluated their performances. Of the three architectures investigated, the styleGAN3D approaches provided the best performance according to FID values and observer studies, while the bigGAN3D architecture was superior in an analysis of 3D structure. The use of minibatch discrimination was shown to be crucial for the styleGAN3D architecture in order to avoid excessive mode collapse and inclusion of a latent mapping network was shown to allow styleGAN3D to produce a latent space with semantic meaning. However, despite the promising results, we conclude that none of the investigated methods are currently suitable for the downstream tasks of anomaly detection or image reconstruction. Future directions of research are proposed, including the use of positional encoding or whole-volume generation, to further improve on these results. Overall, the results and analyses presented in this work provide an important baseline for future research into the use of generative models in 3D healthy lung CT patch modelling.}


\acks{This work was funded by the EPSRC grant [EP/P023509/1] and supported by funding from the Wellcome/EPSRC Centre for Medical Engineering [WT203148/Z/16/Z]; the National Institute for Health Research (NIHR) Biomedical Research Centre based at Guy's and St Thomas' NHS Foundation Trust and King's College London; and the UKRI London Medical Imaging and Artificial Intelligence Centre for Value Based Healthcare.}

%
\ethics{The work follows appropriate ethical standards in conducting research and writing the manuscript, following all applicable laws and regulations regarding treatment of animals or human subjects.}

\coi{There are no conflicts of interest to declare.}

\bibliography{refs}

\begin{thebibliography}{52}
\providecommand{\natexlab}[1]{#1}
\providecommand{\url}[1]{\texttt{#1}}
\expandafter\ifx\csname urlstyle\endcsname\relax
  \providecommand{\doi}[1]{doi: #1}\else
  \providecommand{\doi}{doi: \begingroup \urlstyle{rm}\Url}\fi

\bibitem[Arjovsky et~al.(2017)Arjovsky, Chintala, and Bottou]{Arjovsky2017}
Martin Arjovsky, Soumith Chintala, and L\'{e}on Bottou.
\newblock {Wasserstein Generative Adversarial Networks}.
\newblock In Doina Precup and Yee~Whye Teh, editors, \emph{Proc. 34th Int.
  Conf. Mach. Learn.}, volume~70 of \emph{Proceedings of Machine Learning
  Research}, pages 214--223. PMLR, 2017.

\bibitem[Barile et~al.(2021)Barile, Marzullo, Stamile, Durand-Dubief, and
  Sappey-Marinier]{Barile2021}
Berardino Barile, Aldo Marzullo, Claudio Stamile, Fran\c{c}oise Durand-Dubief,
  and Dominique Sappey-Marinier.
\newblock {Data augmentation using generative adversarial neural networks on
  brain structural connectivity in multiple sclerosis}.
\newblock \emph{Comput. Methods Programs Biomed.}, 206:\penalty0 106113, 2021.

\bibitem[Brock et~al.(2019)Brock, Donahue, and Simonyan]{Brock2019}
Andrew Brock, Jeff Donahue, and Karen Simonyan.
\newblock {Large scale GAN training for high fidelity natural image synthesis}.
\newblock In \emph{7th Int. Conf. Learn. Represent. ICLR 2019}, 2019.

\bibitem[Bu et~al.(2021)Bu, Yang, Jiang, Zhang, Zhang, and Wei]{Bu2021}
Tian Bu, Zhiyong Yang, Shan Jiang, Guobin Zhang, Hongyun Zhang, and Lin Wei.
\newblock {3D conditional generative adversarial network-based synthetic
  medical image augmentation for lung nodule detection}.
\newblock \emph{Int. J. Imaging Syst. Technol.}, 31\penalty0 (2):\penalty0
  670--681, 2021.

\bibitem[Chen et~al.(2019)Chen, Ma, and Zheng]{Chen2019}
Sihong Chen, Kai Ma, and Yefeng Zheng.
\newblock {Med3D: Transfer Learning for 3D Medical Image Analysis}, 2019.
\newblock URL \url{https://arxiv.org/abs/1904.00625}.

\bibitem[Dong et~al.(2019)Dong, Lei, Wang, Thomas, Tang, Curran, Liu, and
  Yang]{Dong2019}
Xue Dong, Yang Lei, Tonghe Wang, Matthew Thomas, Leonardo Tang, Walter~J
  Curran, Tian Liu, and Xiaofeng Yang.
\newblock {Automatic multiorgan segmentation in thorax CT images using
  U-net-GAN}.
\newblock \emph{Med. Phys.}, 46\penalty0 (5):\penalty0 2157--2168, 2019.

\bibitem[Dosovitskiy et~al.(2021)Dosovitskiy, Beyer, Kolesnikov, Weissenborn,
  Zhai, Unterthiner, Dehghani, Minderer, Heigold, Gelly, Uszkoreit, and
  Houlsby]{Dosovitskiy2021}
Alexey Dosovitskiy, Lucas Beyer, Alexander Kolesnikov, Dirk Weissenborn,
  Xiaohua Zhai, Thomas Unterthiner, Mostafa Dehghani, Matthias Minderer, Georg
  Heigold, Sylvain Gelly, Jakob Uszkoreit, and Neil Houlsby.
\newblock {An Image is Worth 16x16 Words: Transformers for Image Recognition at
  Scale}.
\newblock In \emph{International Conference on Learning Representations}, 2021.

\bibitem[Emami et~al.(2018)Emami, Dong, Nejad-Davarani, and
  Glide-Hurst]{Emami2018}
Hajar Emami, Ming Dong, Siamak~P Nejad-Davarani, and Carri~K Glide-Hurst.
\newblock {Generating synthetic CTs from magnetic resonance images using
  generative adversarial networks}.
\newblock \emph{Med. Phys.}, 45\penalty0 (8):\penalty0 3627--3636, 2018.

\bibitem[Frid-Adar et~al.(2018)Frid-Adar, Diamant, Klang, Amitai, Goldberger,
  and Greenspan]{Frid-Adar2018}
Maayan Frid-Adar, Idit Diamant, Eyal Klang, Michal Amitai, Jacob Goldberger,
  and Hayit Greenspan.
\newblock {GAN-based synthetic medical image augmentation for increased CNN
  performance in liver lesion classification}.
\newblock \emph{Neurocomputing}, 321:\penalty0 321--331, 2018.

\bibitem[Goodfellow et~al.(2014)Goodfellow, Pouget-Abadie, Mirza, Xu,
  Warde-Farley, Ozair, Courville, and Bengio]{Goodfellow2014}
Ian Goodfellow, Jean Pouget-Abadie, Mehdi Mirza, Bing Xu, David Warde-Farley,
  Sherjil Ozair, Aaron Courville, and Yoshua Bengio.
\newblock {Generative Adversarial Nets}.
\newblock In Z~Ghahramani, M~Welling, C~Cortes, N~Lawrence, and K~Q Weinberger,
  editors, \emph{Adv. Neural Inf. Process. Syst.}, volume~27. Curran
  Associates, Inc., 2014.

\bibitem[Gulrajani et~al.(2017)Gulrajani, Ahmed, Arjovsky, Dumoulin, and
  Courville]{Gulrajani2017}
Ishaan Gulrajani, Faruk Ahmed, Martin Arjovsky, Vincent Dumoulin, and Aaron~C
  Courville.
\newblock {Improved Training of Wasserstein GANs}.
\newblock In I~Guyon, U~V Luxburg, S~Bengio, H~Wallach, R~Fergus,
  S~Vishwanathan, and R~Garnett, editors, \emph{Adv. Neural Inf. Process.
  Syst.}, volume~30. Curran Associates, Inc., 2017.

\bibitem[Heusel et~al.(2017)Heusel, Ramsauer, Unterthiner, Nessler, and
  Hochreiter]{Heusel2017}
Martin Heusel, Hubert Ramsauer, Thomas Unterthiner, Bernhard Nessler, and Sepp
  Hochreiter.
\newblock {GANs Trained by a Two Time-Scale Update Rule Converge to a Local
  Nash Equilibrium}.
\newblock In I~Guyon, U~V Luxburg, S~Bengio, H~Wallach, R~Fergus,
  S~Vishwanathan, and R~Garnett, editors, \emph{Adv. Neural Inf. Process.
  Syst.}, volume~30. Curran Associates, Inc., 2017.

\bibitem[Hong et~al.(2021)Hong, Marinescu, Dalca, Bonkhoff, Bretzner, Rost, and
  Golland]{Hong2021}
Sungmin Hong, Razvan Marinescu, Adrian~V. Dalca, Anna~K. Bonkhoff, Martin
  Bretzner, Natalia~S. Rost, and Polina Golland.
\newblock {3D-StyleGAN: A Style-Based Generative Adversarial Network for
  Generative Modeling of Three-Dimensional Medical Images}.
\newblock In Sandy Engelhardt, Ilkay Oksuz, Dajiang Zhu, Yixuan Yuan, Anirban
  Mukhopadhyay, Nicholas Heller, Sharon~Xiaolei Huang, Hien Nguyen, Raphael
  Sznitman, and Yuan Xue, editors, \emph{Deep Generative Models, and Data
  Augmentation, Labelling, and Imperfections}, pages 24--34, Cham, 2021.
  Springer International Publishing.

\bibitem[Huang and Belongie(2017)]{Huang2017}
Xun Huang and Serge Belongie.
\newblock {Arbitrary Style Transfer in Real-time with Adaptive Instance
  Normalization}.
\newblock In \emph{Proceedings of the International Conference on Computer
  Vision (ICCV)}, volume 2017-Octob, pages 1510--1519. Institute of Electrical
  and Electronics Engineers Inc., 2017.

\bibitem[Ioffe and Szegedy(2015)]{Ioffe2015}
Sergey Ioffe and Christian Szegedy.
\newblock {Batch Normalization: Accelerating Deep Network Training by Reducing
  Internal Covariate Shift}.
\newblock In Francis Bach and David Blei, editors, \emph{Proc. 32nd Int. Conf.
  Mach. Learn.}, volume~37 of \emph{Proceedings of Machine Learning Research},
  pages 448--456, Lille, France, 2015. PMLR.

\bibitem[Isola et~al.(2017)Isola, Zhu, Zhou, and Efros]{Isola2018}
Phillip Isola, Jun-Yan Zhu, Tinghui Zhou, and Alexei~A. Efros.
\newblock Image-to-image translation with conditional adversarial networks.
\newblock In \emph{Proceedings of the IEEE Conference on Computer Vision and
  Pattern Recognition (CVPR)}, 2017.

\bibitem[Jin et~al.(2018)Jin, Xu, Tang, Harrison, and Mollura]{Jin2018}
Dakai Jin, Ziyue Xu, Youbao Tang, Adam~P. Harrison, and Daniel~J. Mollura.
\newblock {CT-Realistic Lung Nodule Simulation from 3D Conditional Generative
  Adversarial Networks for Robust Lung Segmentation}.
\newblock In Alejandro~F. Frangi, Julia~A. Schnabel, Christos Davatzikos,
  Carlos Alberola-L\'{o}pez, and Gabor Fichtinger, editors, \emph{Medical Image
  Computing and Computer Assisted Intervention -- MICCAI 2018}, pages 732--740,
  Cham, 2018. Springer International Publishing.

\bibitem[Jolicoeur-Martineau(2019)]{Jolicoeur-Martineau2018}
Alexia Jolicoeur-Martineau.
\newblock The relativistic discriminator: a key element missing from standard
  {GAN}.
\newblock In \emph{International Conference on Learning Representations}, 2019.
\newblock URL \url{https://openreview.net/forum?id=S1erHoR5t7}.

\bibitem[Karras et~al.(2018)Karras, Aila, Laine, and Lehtinen]{Karras2018}
Tero Karras, Timo Aila, Samuli Laine, and Jaakko Lehtinen.
\newblock {Progressive growing of GANs for improved quality, stability, and
  variation}.
\newblock In \emph{6th Int. Conf. Learn. Represent. ICLR 2018 - Conf. Track
  Proc.}, 2018.

\bibitem[Karras et~al.(2019)Karras, Laine, and Aila]{Karras2019}
Tero Karras, Samuli Laine, and Timo Aila.
\newblock {A style-based generator architecture for generative adversarial
  networks}.
\newblock In \emph{Proc. IEEE Comput. Soc. Conf. Comput. Vis. Pattern
  Recognit.}, volume 2019-June, pages 4396--4405, 2019.

\bibitem[Karras et~al.(2020)Karras, Laine, Aittala, Hellsten, Lehtinen, and
  Aila]{Karras2020}
Tero Karras, Samuli Laine, Miika Aittala, Janne Hellsten, Jaakko Lehtinen, and
  Timo Aila.
\newblock {Analyzing and Improving the Image Quality of StyleGAN}.
\newblock In \emph{Proceedings of the IEEE Conference on Computer Vision and
  Pattern Recognition (CVPR)}, pages 8107--8116, 2020.

\bibitem[Karras et~al.(2021)Karras, Aittala, Laine, H\"{a}rk\"{o}nen, Hellsten,
  Lehtinen, and Aila]{Karras2021}
Tero Karras, Miika Aittala, Samuli Laine, Erik H\"{a}rk\"{o}nen, Janne
  Hellsten, Jaakko Lehtinen, and Timo Aila.
\newblock {Alias-Free Generative Adversarial Networks}.
\newblock In M.~Ranzato, A.~Beygelzimer, Y.~Dauphin, P.S. Liang, and J.~Wortman
  Vaughan, editors, \emph{Advances in Neural Information Processing Systems},
  volume~34, pages 852--863. Curran Associates, Inc., 2021.

\bibitem[Kingma and Ba(2015)]{Kingma2015}
Diederik~P. Kingma and Jimmy~Lei Ba.
\newblock {Adam: A method for stochastic optimization}.
\newblock In \emph{3rd Int. Conf. Learn. Represent. ICLR 2015 - Conf. Track
  Proc.}, 2015.

\bibitem[Lample et~al.(2017)Lample, Zeghidour, Usunier, Bordes, DENOYER, and
  Ranzato]{Lample2017}
Guillaume Lample, Neil Zeghidour, Nicolas Usunier, Antoine Bordes, Ludovic
  DENOYER, and Marc\textquotesingle~Aurelio Ranzato.
\newblock {Fader Networks: Manipulating Images by Sliding Attributes}.
\newblock In I.~Guyon, U.~V. Luxburg, S.~Bengio, H.~Wallach, R.~Fergus,
  S.~Vishwanathan, and R.~Garnett, editors, \emph{Advances in Neural
  Information Processing Systems}, volume~30. Curran Associates, Inc., 2017.

\bibitem[Lim and Ye(2017)]{Lim2017}
Jae~Hyun Lim and Jong~Chul Ye.
\newblock {Geometric GAN}.
\newblock \emph{arXiv}, page 1705.02894, 2017.

\bibitem[Lin et~al.(2019)Lin, Chang, Chen, Juan, Wei, and Chen]{Lin2020}
Chieh~Hubert Lin, Chia-Che Chang, Yu-Sheng Chen, Da-Cheng Juan, Wei Wei, and
  Hwann-Tzong Chen.
\newblock {COCO-GAN: Generation by Parts via Conditional Coordinating}.
\newblock In \emph{Proceedings of the International Conference on Computer
  Vision (ICCV)}, 2019.

\bibitem[Liu et~al.(2018)Liu, Gibson, Grbic, Xu, Setio, Yang, Georgescu, and
  Comaniciu]{Liu2018}
Siqi Liu, Eli Gibson, Sasa Grbic, Zhoubing Xu, Arnaud Arindra~Adiyoso Setio,
  Jie Yang, Bogdan Georgescu, and Dorin Comaniciu.
\newblock {Decompose to manipulate: Manipulable Object Synthesis in 3D Medical
  Images with Structured Image Decomposition}.
\newblock \emph{arXiv}, page 1812.01737, 2018.

\bibitem[Mao et~al.(2017)Mao, Li, Xie, Lau, Wang, and Smolley]{Mao2017}
Xudong Mao, Qing Li, Haoran Xie, Raymond Y~K Lau, Zhen Wang, and Stephen~Paul
  Smolley.
\newblock {Least Squares Generative Adversarial Networks}.
\newblock In \emph{Proceedings of the International Conference on Computer
  Vision (ICCV)}, pages 2813--2821, 2017.

\bibitem[McInnes et~al.(2020)McInnes, Healy, and Melville]{McInnes2020}
Leland McInnes, John Healy, and James Melville.
\newblock {UMAP: Uniform Manifold Approximation and Projection for Dimension
  Reduction}.
\newblock \emph{arXiv}, page 1802.03426, 2020.

\bibitem[Miyato et~al.(2018)Miyato, Kataoka, Koyama, and Yoshida]{Miyato2018}
Takeru Miyato, Toshiki Kataoka, Masanori Koyama, and Yuichi Yoshida.
\newblock {Spectral normalization for generative adversarial networks}.
\newblock In \emph{6th Int. Conf. Learn. Represent. ICLR 2018 - Conf. Track
  Proc.}, 2018.

\bibitem[Montalt-Tordera et~al.(2021)Montalt-Tordera, Muthurangu, Hauptmann,
  and Steeden]{Montalt-Tordera2021}
Javier Montalt-Tordera, Vivek Muthurangu, Andreas Hauptmann, and Jennifer~Anne
  Steeden.
\newblock {Machine learning in Magnetic Resonance Imaging: Image
  reconstruction}.
\newblock \emph{Phys. Medica}, 83:\penalty0 79--87, 2021.

\bibitem[Perarnau et~al.(2016)Perarnau, van~de Weijer, Raducanu, and
  \'{A}lvarez]{Perarnau2016}
Guim Perarnau, Joost van~de Weijer, Bogdan Raducanu, and Jose~M \'{A}lvarez.
\newblock {Invertible Conditional GANs for image editing}.
\newblock \emph{arXiv}, page 1611.06355, 2016.

\bibitem[Pesaranghader et~al.(2021)Pesaranghader, Wang, and
  Havaei]{Pesaranghader2021}
Ahmad Pesaranghader, Yiping Wang, and Mohammad Havaei.
\newblock {CT-SGAN: Computed Tomography Synthesis GAN}.
\newblock In Sandy Engelhardt, Ilkay Oksuz, Dajiang Zhu, Yixuan Yuan, Anirban
  Mukhopadhyay, Nicholas Heller, Sharon~Xiaolei Huang, Hien Nguyen, Raphael
  Sznitman, and Yuan Xue, editors, \emph{Deep Generative Models, and Data
  Augmentation, Labelling, and Imperfections}, pages 67--79, Cham, 2021.
  Springer International Publishing.

\bibitem[Quiros et~al.(2021)Quiros, Murray-Smith, and Yuan]{Quiros2021}
Adalberto~Claudio Quiros, Roderick Murray-Smith, and Ke~Yuan.
\newblock {PathologyGAN: Learning deep representations of cancer tissue}.
\newblock \emph{Machine Learning for Biomedical Imaging}, 1, 2021.

\bibitem[Radford et~al.(2016)Radford, Metz, and Chintala]{Radford2016}
Alec Radford, Luke Metz, and Soumith Chintala.
\newblock {Unsupervised representation learning with deep convolutional
  generative adversarial networks}.
\newblock In \emph{4th Int. Conf. Learn. Represent. ICLR 2016 - Conf. Track
  Proc.} International Conference on Learning Representations, ICLR, 2016.

\bibitem[Ravishankar et~al.(2020)Ravishankar, Ye, and Fessler]{Ravishankar2020}
Saiprasad Ravishankar, Jong~Chul Ye, and Jeffrey~A Fessler.
\newblock {Image Reconstruction: From Sparsity to Data-Adaptive Methods and
  Machine Learning}.
\newblock \emph{Proc. IEEE}, 108\penalty0 (1):\penalty0 86--109, 2020.

\bibitem[Razavi et~al.(2019)Razavi, van~den Oord, and Vinyals]{Razavi2019}
Ali Razavi, Aaron van~den Oord, and Oriol Vinyals.
\newblock Generating diverse high-fidelity images with vq-vae-2.
\newblock In H.~Wallach, H.~Larochelle, A.~Beygelzimer, F.~d\textquotesingle
  Alch\'{e}-Buc, E.~Fox, and R.~Garnett, editors, \emph{Advances in Neural
  Information Processing Systems}, volume~32. Curran Associates, Inc., 2019.
\newblock URL
  \url{https://proceedings.neurips.cc/paper/2019/file/5f8e2fa1718d1bbcadf1cd9c7a54fb8c-Paper.pdf}.

\bibitem[Reader et~al.(2020)Reader, Corda, Mehranian, da~Costa-Luis, Ellis, and
  Schnabel]{Reader2020}
Andrew~J. Reader, Guillaume Corda, Abolfazl Mehranian, Casper da~Costa-Luis,
  Sam Ellis, and Julia~A. Schnabel.
\newblock {Deep Learning for PET Image Reconstruction}.
\newblock \emph{IEEE Trans. Radiat. Plasma Med. Sci.}, 5\penalty0 (1):\penalty0
  1--25, 2020.

\bibitem[Salimans et~al.(2016)Salimans, Goodfellow, Zaremba, Cheung, Radford,
  and Chen]{Salimans2016}
Tim Salimans, Ian Goodfellow, Wojciech Zaremba, Vicki Cheung, Alec Radford, and
  Xi~Chen.
\newblock {Improved Techniques for Training GANs}.
\newblock \emph{arXiv}, page 1606.03498v1, 2016.

\bibitem[Schlegl et~al.(2019)Schlegl, Seeb\"{o}ck, Waldstein, Langs, and
  Schmidt-Erfurth]{Schlegl2019}
Thomas Schlegl, Philipp Seeb\"{o}ck, Sebastian~M. Waldstein, Georg Langs, and
  Ursula Schmidt-Erfurth.
\newblock {f-AnoGAN: Fast unsupervised anomaly detection with generative
  adversarial networks}.
\newblock \emph{Med. Image Anal.}, 54:\penalty0 30--44, 2019.

\bibitem[Setio et~al.(2017)Setio, Traverso, de~Bel, Berens, van~den Bogaard,
  Cerello, Chen, Dou, Fantacci, Geurts, van~der Gugten, Heng, Jansen, de~Kaste,
  Kotov, Lin, Manders, S\'{o}\~{n}ora Mengana, Garc\'{i}­a-Naranjo,
  Papavasileiou, Prokop, Saletta, Schaefer-Prokop, Scholten, Scholten, Snoeren,
  Torres, Vandemeulebroucke, Walasek, Zuidhof, van Ginneken, and
  Jacobs]{Setio2017}
Arnaud Arindra~Adiyoso Setio, Alberto Traverso, Thomas de~Bel, Moira~S.N.
  Berens, Cas van~den Bogaard, Piergiorgio Cerello, Hao Chen, Qi~Dou,
  Maria~Evelina Fantacci, Bram Geurts, Robbert van~der Gugten, Pheng~Ann Heng,
  Bart Jansen, Michael~M.J. de~Kaste, Valentin Kotov, Jack Yu-Hung Lin,
  Jeroen~T.M.C. Manders, Alexander S\'{o}\~{n}ora Mengana, Juan~Carlos
  Garc\'{i}­a-Naranjo, Evgenia Papavasileiou, Mathias Prokop, Marco Saletta,
  Cornelia~M Schaefer-Prokop, Ernst~T. Scholten, Luuk Scholten, Miranda~M.
  Snoeren, Ernesto~Lopez Torres, Jef Vandemeulebroucke, Nicole Walasek,
  Guido~C.A. Zuidhof, Bram van Ginneken, and Colin Jacobs.
\newblock {Validation, comparison, and combination of algorithms for automatic
  detection of pulmonary nodules in computed tomography images: The LUNA16
  challenge}.
\newblock \emph{Med. Image Anal.}, 42:\penalty0 1--13, 2017.

\bibitem[Sinha et~al.(2020)Sinha, Zhao, Goyal, Raffel, and Odena]{Sinha2020}
Samarth Sinha, Zhengli Zhao, Anirudh Goyal, Colin Raffel, and Augustus Odena.
\newblock {Top-k Training of GANs: Improving GAN Performance by Throwing Away
  Bad Samples}.
\newblock In H.~Larochelle, M.~Ranzato, R.~Hadsell, M.~F. Balcan, and H.~Lin,
  editors, \emph{Advances in Neural Information Processing Systems}, volume~33,
  pages 14638--14649. Curran Associates, Inc., 2020.

\bibitem[Suh et~al.(2021)Suh, Cheon, Chang, Lee, and Lee]{Suh2021}
Sungho Suh, Sojeong Cheon, Dong-Jin Chang, Deukhee Lee, and Yong~Oh Lee.
\newblock {Sequential Lung Nodule Synthesis Using Attribute-Guided Generative
  Adversarial Networks}.
\newblock In Marleen de~Bruijne, Philippe~C. Cattin, St\'{e}phane Cotin,
  Nicolas Padoy, Stefanie Speidel, Yefeng Zheng, and Caroline Essert, editors,
  \emph{Medical Image Computing and Computer Assisted Intervention -- MICCAI
  2021}, pages 402--411, Cham, 2021. Springer International Publishing.

\bibitem[Sun et~al.(2020)Sun, Chen, Xu, Gong, Yu, and Batmanghelich]{Sun2020}
Li~Sun, Junxiang Chen, Yanwu Xu, Mingming Gong, Ke~Yu, and Kayhan
  Batmanghelich.
\newblock {Hierarchical Amortized Training for Memory-efficient High Resolution
  3D GAN }.
\newblock In \emph{MedNeurIPS 2020}, 2020.

\bibitem[Uzunova et~al.(2020)Uzunova, Ehrhardt, and Handels]{Uzunova2020}
Hristina Uzunova, Jan Ehrhardt, and Heinz Handels.
\newblock {Memory-efficient {GAN}-based domain translation of high resolution
  3D medical images}.
\newblock \emph{Computerized Medical Imaging and Graphics}, 86:\penalty0
  101801, 2020.

\bibitem[van~den Oord et~al.(2017)van~den Oord, Vinyals, and
  Kavukcuoglu]{Oord2017}
Aaron van~den Oord, Oriol Vinyals, and Koray Kavukcuoglu.
\newblock Neural discrete representation learning.
\newblock In I.~Guyon, U.~Von Luxburg, S.~Bengio, H.~Wallach, R.~Fergus,
  S.~Vishwanathan, and R.~Garnett, editors, \emph{Advances in Neural
  Information Processing Systems}, volume~30. Curran Associates, Inc., 2017.
\newblock URL
  \url{https://proceedings.neurips.cc/paper/2017/file/7a98af17e63a0ac09ce2e96d03992fbc-Paper.pdf}.

\bibitem[Wang et~al.(2021)Wang, Zhang, Zhang, Gao, Huang, Wang, Zhang, Yang,
  and Liu]{Wang2021}
Qiuli Wang, Xiaohong Zhang, Wei Zhang, Mingchen Gao, Sheng Huang, Jian Wang,
  Jiuquan Zhang, Dan Yang, and Chen Liu.
\newblock {Realistic Lung Nodule Synthesis With Multi-Target Co-Guided
  Adversarial Mechanism}.
\newblock \emph{IEEE Transactions on Medical Imaging}, 40\penalty0
  (9):\penalty0 2343--2353, 2021.

\bibitem[Xu et~al.(2019)Xu, Wang, Shin, Roth, Yang, Milletari, Zhang, and
  Xu]{Xu2019}
Ziyue Xu, Xiaosong Wang, Hoo-Chang Shin, Holger Roth, Dong Yang, Fausto
  Milletari, Ling Zhang, and Daguang Xu.
\newblock {Tunable CT Lung Nodule Synthesis Conditioned on Background Image and
  Semantic Features}.
\newblock In Ninon Burgos, Ali Gooya, and David Svoboda, editors,
  \emph{Simulation and Synthesis in Medical Imaging}, pages 62--70, Cham, 2019.
  Springer International Publishing.

\bibitem[Yang et~al.(2019)Yang, Liu, Grbic, Setio, Xu, Gibson, Chabin,
  Georgescu, Laine, and Comaniciu]{Yang2019}
Jie Yang, Siqi Liu, Sasa Grbic, Arnaud Arindra~Adiyoso Setio, Zhoubing Xu, Eli
  Gibson, Guillaume Chabin, Bogdan Georgescu, Andrew~F. Laine, and Dorin
  Comaniciu.
\newblock {Class-Aware Adversarial Lung Nodule Synthesis In CT Images}.
\newblock In \emph{2019 IEEE 16th International Symposium on Biomedical Imaging
  (ISBI 2019)}, pages 1348--1352, 2019.

\bibitem[Yi et~al.(2019)Yi, Walia, and Babyn]{Yi2019}
Xin Yi, Ekta Walia, and Paul Babyn.
\newblock {Generative adversarial network in medical imaging: A review}.
\newblock \emph{Med. Image Anal.}, 58:\penalty0 101552, 2019.

\bibitem[Zhang et~al.(2019)Zhang, Goodfellow, Metaxas, and Odena]{Zhang2019}
Han Zhang, Ian Goodfellow, Dimitris Metaxas, and Augustus Odena.
\newblock {Self-Attention Generative Adversarial Networks}.
\newblock In Kamalika Chaudhuri and Ruslan Salakhutdinov, editors,
  \emph{Proceedings of the 36th International Conference on Machine Learning},
  volume~97 of \emph{Proceedings of Machine Learning Research}, pages
  7354--7363. PMLR, 2019.

\bibitem[Zhu et~al.(2017)Zhu, Park, Isola, and Efros]{Zhu2017}
Jun~Yan Zhu, Taesung Park, Phillip Isola, and Alexei~A. Efros.
\newblock {Unpaired Image-to-Image Translation Using Cycle-Consistent
  Adversarial Networks}.
\newblock In \emph{Proceedings of the International Conference on Computer
  Vision (ICCV)}, volume 2017-Octob, pages 2242--2251, 2017.

\end{thebibliography}

\newpage
\appendix 

\renewcommand{\thesection}{\Alph{section}}

\section{DCGAN3D generator architecture}
\label{sec:dcgan_archi}

\setcounter{table}{0}
\setcounter{figure}{0}
\renewcommand{\thetable}{A\arabic{table}}
\renewcommand{\thefigure}{A\arabic{figure}}

\begin{table}[h]
\footnotesize
\begin{center}
\begin{tabular}{c}
\toprule
\textbf{DCGAN3D generator}\\
\midrule
Input: $z \in \mathbb{R}^{512} \sim \mathcal{N}(0,1)$\\
\midrule
ConvTranspose3d $4\times4\times4$, stride 1, pad 0, no bias, 512$\rightarrow$512\\
\midrule
BatchNorm3d + ReLU\\
\midrule
ConvTranspose3d $2\times4\times4$, stride 2, pad $2\times 1 \times 1$, no bias, 512$\rightarrow$256\\
\midrule
BatchNorm3d + ReLU\\
\midrule
ConvTranspose3d $4\times4\times4$, stride 2, pad 1, no bias, 256$\rightarrow$128\\
\midrule
BatchNorm3d + ReLU\\
\midrule
ConvTranspose3d $4\times4\times4$, stride 2, pad 1, no bias, 128$\rightarrow$64\\
\midrule
BatchNorm3d + ReLU\\
\midrule
ConvTranspose3d $4\times4\times4$, stride 2, pad 1, no bias, 64$\rightarrow$1\\
\midrule
Tanh\\
\midrule
Output: $x_f \in \mathbb{R}^{32 \times 64 \times 64} \sim \mathbb{Q}$\\
\bottomrule
\end{tabular}
\end{center}
\caption{Architecture of the 3D DCGAN method used in this investigation.}
\label{table:DCGAN_gen_archi}
\end{table}

\clearpage
\section{styleGAN3D generator architecture}
\label{sec:stylegan_archi}

\begin{table}[h]
\footnotesize
\begin{center}
\begin{tabular}{c}
\toprule
\textbf{styleGAN3D generator: latent mapping network}\\
\midrule
Input: $z \in \mathbb{R}^{512} \sim \mathcal{N}(0,1)$\\
\midrule
FC, 512$\rightarrow$ 512 \\
\midrule
Leaky ReLU 0.2 \\
\midrule
FC, 512$\rightarrow$ 512 \\
\midrule
Leaky ReLU 0.2 \\
\midrule
FC 512$\rightarrow$ 512 \\
\midrule
Leaky ReLU 0.2 \\
\midrule
FC, 512$\rightarrow$ 512 \\
\midrule
Leaky ReLU 0.2 \\
\midrule
FC, 512$\rightarrow$ 512 \\
\midrule
Leaky ReLU 0.2 \\
\midrule
FC, 512$\rightarrow$ 512 \\
\midrule
Leaky ReLU 0.2 \\
\midrule
FC, 512$\rightarrow$ 512 \\
\midrule
Leaky ReLU 0.2 \\
\midrule
FC, 512$\rightarrow$ 512 \\
\midrule
Leaky ReLU 0.2 \\
\midrule
Output: $w \in \mathbb{R}^{512}$\\
\bottomrule
\end{tabular}
\end{center}
\caption{Architecture of the latent mapping network in the styleGAN3D generator. FC = fully connected layer.}
\label{table:styleGAN_gen_latent_mapping}
\end{table}

\begin{table}[h]
\footnotesize
\begin{center}
\begin{tabular}{c}
\toprule
\textbf{styleGAN3D generator: synthesis network}\\
\midrule
Input: $w \in \mathbb{R}^{512}$, $\textrm{Const.} \in \mathbb{R}^{1 \times 2 \times 2}$ \\
\midrule
AdaIN, Conv3d $3 \times 3 \times 3$, stride 1, pad 1, no bias, 512$\rightarrow512$ \\
\midrule
Leaky ReLU 0.2 \\
\midrule
AdaIN, $2 \times$ nearest-neighbour upsample, Conv3d $3 \times 3 \times 3$, stride 1, pad 1, no bias, 512$\rightarrow256$ \\
\midrule
Leaky ReLU 0.2 \\
\midrule
AdaIN, Conv3d $3 \times 3 \times 3$, stride 1, pad 1, no bias, 256$\rightarrow256$ \\
\midrule
Leaky ReLU 0.2 \\
\midrule
AdaIN, $2 \times$ nearest-neighbour upsample, Conv3d $3 \times 3 \times 3$, stride 1, pad 1, no bias, 256$\rightarrow128$ \\
\midrule
Leaky ReLU 0.2 \\
\midrule
AdaIN, Conv3d $3 \times 3 \times 3$, stride 1, pad 1, no bias, 128$\rightarrow128$ \\
\midrule
Leaky ReLU 0.2 \\
\midrule
AdaIN, $2 \times$ nearest-neighbour upsample, Conv3d $3 \times 3 \times 3$, stride 1, pad 1, no bias, 128$\rightarrow64$ \\
\midrule
Leaky ReLU 0.2 \\
\midrule
AdaIN, Conv3d $3 \times 3 \times 3$, stride 1, pad 1, no bias, 64$\rightarrow64$ \\
\midrule
AdaIN, $2 \times$ nearest-neighbour upsample, Conv3d $3 \times 3 \times 3$, stride 1, pad 1, no bias, 64$\rightarrow32$ \\
\midrule
Leaky ReLU 0.2 \\
\midrule
AdaIN, Conv3d $3 \times 3 \times 3$, stride 1, pad 1, no bias, 32$\rightarrow32$ \\
\midrule
Leaky ReLU 0.2 \\
\midrule
AdaIN, $2 \times$ nearest-neighbour upsample, Conv3d $3 \times 3 \times 3$, stride 1, pad 1, no bias, 32$\rightarrow16$ \\
\midrule
Leaky ReLU 0.2 \\
\midrule
AdaIN, Conv3d $3 \times 3 \times 3$, stride 1, pad 1, no bias, 16$\rightarrow1$ \\
\midrule
Tanh \\
\midrule
Output: $x_f \in \mathbb{R}^{32 \times 64 \times 64} \sim \mathbb{Q}$\\
\bottomrule
\end{tabular}
\end{center}
\caption{Architecture of the synthesis network in the styleGAN3D generator.}
\label{table:styleGAN_gen_synthesis}
\end{table}

\clearpage
\section{\textblue{bigGAN3D generator architecture}}
\label{sec:biggan_archi}

\begin{table}[h]
\footnotesize
\begin{center}
\begin{tabular}{c}
\toprule
\textbf{bigeGAN3D generator}\\
\midrule
Input: $z \in \mathbb{R}^{512} \sim \mathcal{N}(0,1)$ \\
\midrule
SNFC, 512 $\rightarrow$ 6144\\
\midrule
Reshape \\
\midrule
BatchNorm3d + ReLU \\
\midrule
$(1 \times 2 \times 2) \times$ nearest neighbour upsample, SNConv3d $3 \times 3 \times 3$, stride 1, pad 1, bias, 96 $\rightarrow$ 96 \\
\midrule
BatchNorm3d + ReLU \\
\midrule
SNConv3d $3 \times 3 \times 3$, stride 1, pad 1, bias, 96 $\rightarrow$ 96 \\
\midrule
SNConv3d $1 \times 1 \times 1$, stride 1, pad 0, bias, 96 $\rightarrow$ 96 \\
\midrule
SelfAttention3D \\
\midrule
BatchNorm3d + ReLU \\
\midrule
$2 \times$ nearest neighbour upsample, SNConv3d $3 \times 3 \times 3$, stride 1, pad 1, bias, 96 $\rightarrow$ 48 \\
\midrule
BatchNorm3d + ReLU \\
\midrule
SNConv3d $3 \times 3 \times 3$, stride 1, pad 1, bias, 48 $\rightarrow$ 48 \\
\midrule
SNConv3d $1 \times 1 \times 1$, stride 1, pad 0, bias, 48 $\rightarrow$ 48 \\
\midrule
BatchNorm3d + ReLU \\
\midrule
$2 \times$ nearest neighbour upsample, SNConv3d $3 \times 3 \times 3$, stride 1, pad 1, bias, 48 $\rightarrow$ 24 \\
\midrule
BatchNorm3d + ReLU \\
\midrule
SNConv3d $3 \times 3 \times 3$, stride 1, pad 1, bias, 24 $\rightarrow$ 24 \\
\midrule
SNConv3d $1 \times 1 \times 1$, stride 1, pad 0, bias, 24 $\rightarrow$ 24 \\
\midrule
BatchNorm3d + ReLU \\
\midrule
$2 \times$ nearest neighbour upsample, SNConv3d $3 \times 3 \times 3$, stride 1, pad 1, bias, 24 $\rightarrow$ 12 \\
\midrule
BatchNorm3d + ReLU \\
\midrule
SNConv3d $3 \times 3 \times 3$, stride 1, pad 1, bias, 12 $\rightarrow$ 12 \\
\midrule
SNConv3d $1 \times 1 \times 1$, stride 1, pad 0, bias, 12 $\rightarrow$ 12 \\
\midrule
BatchNorm3d + ReLU \\
\midrule
SNConv3d $3 \times 3 \times 3$, stride 1, pad 1, bias, 12 $\rightarrow$ 1 \\
\midrule
Tanh \\
\midrule
Output: $x_f \in \mathbb{R}^{32 \times 64 \times 64} \sim \mathbb{Q}$\\
\bottomrule
\end{tabular}
\end{center}
\caption{\textblue{Architecture of the 3D bigGAN generator. SNFC and SNConv refer to spectrally normalised fully connected and spectrally normalised convolution layers respectively.}}
\label{table:styleGAN_gen_synthesis}
\end{table}

\clearpage
\section{\textblue{Minibatch Discrimination and Increasing the Effective Batch Size}}
\label{sec:mdmin}

\subsection*{\textblue{Minibatch Discrimination: MDmin}}

\textblue{The following approach to MD was adopted, calculating the L$_1$ distance, $d_{ij}$, between two images $x_i$ and $x_j$ within a minibatch as:
\begin{equation}
    d_{ij} = \left\| D_{l}\left(x_i\right) - D_{l}\left(x_j \right) \right\|_1
\end{equation}
where $D_{l}\left(x\right)$ denotes the features of the discriminator network after layer $l$ with $x$ as the input image. In the traditional minibatch discrimination approach \citep{Salimans2016}, these distances (or some transformation of them) are then either summed or averaged over the minibatch to get a per-image score, i.e. $\textrm{MD}_i \propto \sum_{j}^{N} d_{ij}$, which is used as input for the subsequent discriminator layer.

We note that taking the sum of $d_{ij}$ allows partial mode collapse to occur if one mode is sufficiently far from the rest of the samples. To overcome this, we propose to take instead the \emph{minimum} within-batch distance for each image, denoted MDmin, as follows: $\textrm{MDmin}_i = \min_j{(d_{ij})}$. This represents the maximum similarity between two samples in the minibatch, providing a consistently clearer signal to the discriminator if these images are mode-collapsed. The scalar values per image, $\textrm{MDmin}_i$, are then expanded to the correct size and appended to the current feature maps, to provide an extra feature to the subsequent layers of the discriminator.}

\subsection*{\textblue{Increased Effective Batch Size: largeEBS}}

\begin{figure*}[h]
\centering
\includegraphics[width=1\textwidth]{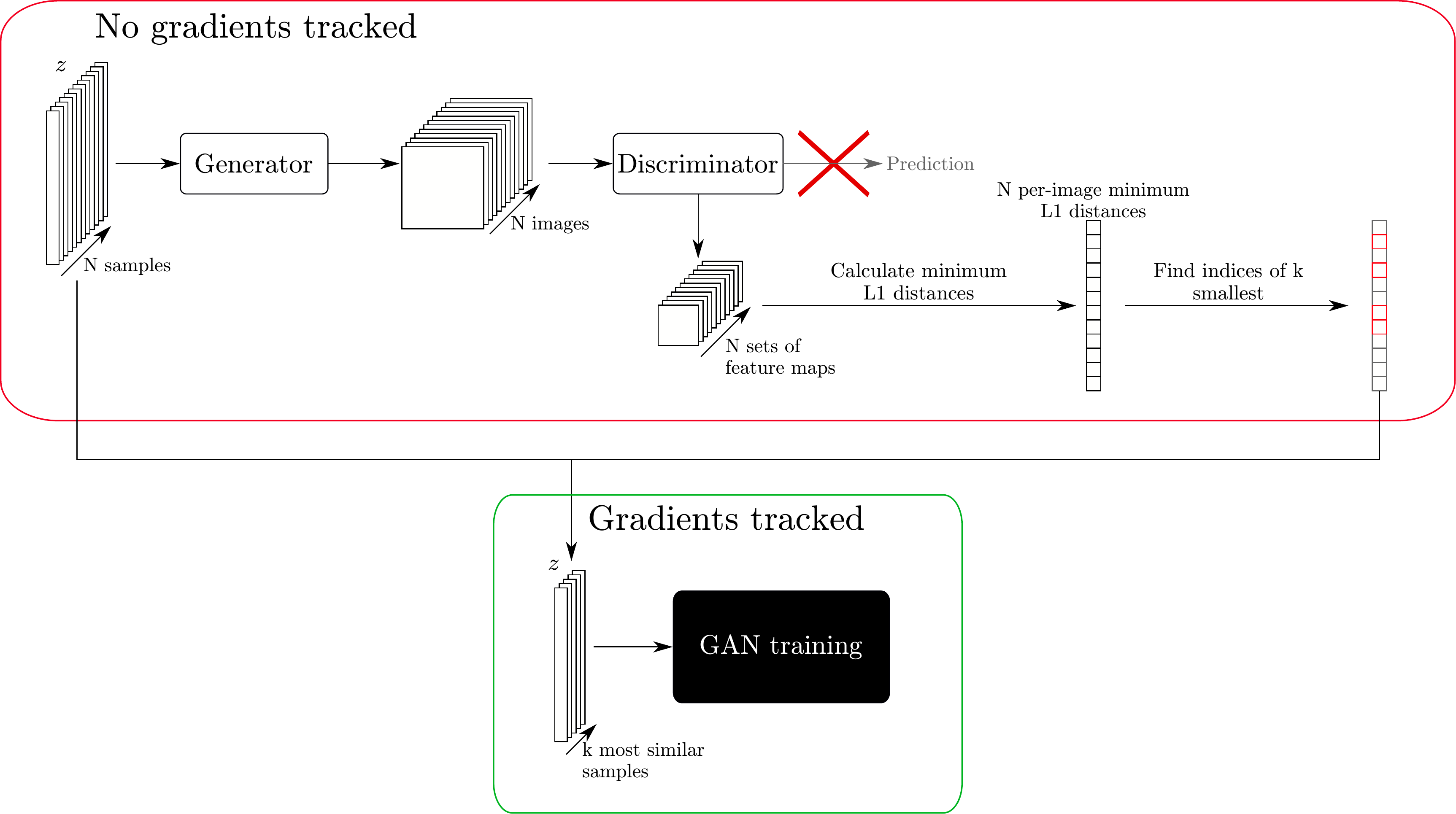}
\caption{\textblue{The proposed large effective batch size (largeEBS) technique. A number, $N$, of latent codes $z$ are sampled from $\mathbb{P}_z$ such that $N$ would not fit in memory for standard training purposes. Without tracking gradients, $z$ are then passed through the generator to produce $N$ generated samples. To detect which images are most mode-collapsed, $L_1$ distances between features from layer $l$ of the discriminator are calculated, and the minimum distances are retained as a measure of degree of mode collapse. The $k$ most mode-collapsed samples are then kept and used for a subsequent training iteration, while the other $N-k$ samples are discarded.}}
\label{fig:largeEBS}
\end{figure*}

\textblue{The MDmin layer described above is able to reduce mode collapse by allowing the discriminator to detect when a minibatch contains similar samples, and feed this information back to the generator. However, mode collapse can clearly still occur, as long as similar samples are unlikely to occur within a minibatch. Size the maximum batch size is bounded by computational limitations (GPU memory), this issue cannot be solved completely by simply increasing the batch size.

We propose to use a heuristic trick to increase the effective batch size, mitigating the effect described above. First, a large batch of $N$ latent codes is sampled and passed through the generator without tracking gradients. This allows a larger batch size than would be normally used in training, since much of the computational load of network training is in the storing of gradients. These generated images are then passed through the discriminator without tracking gradients, and the activation maps at depth $l$ are calculated, similarly as for the MDmin layer. The similarity between all pairs of samples is then calculated, as for the MDmin layer, and the minimum for each sample is taken as a measure of how mode-collapsed that particular sample is within the larger minibatch. The $k$ most similar images are noted, and the corresponding samples are then used for the subsequent training iteration, ensuring that the networks see the most mode-collapsed samples for training purposes. Figure \ref{fig:largeEBS} shows this process schematically. Note that with this method, the upper limit to the effective batch size is no longer dependent on memory, since each pair of images can in theory be loaded to memory as required and do not need to be stored otherwise. Real samples are still selected entirely at random; choosing the $k$ most similar samples from a larger batch of real images was seen to degrade performance. Since the learned discriminator representations will not be relevant early in training, we train the networks for $5$ epochs as normal before increasing the effective batch size thereafter. Similar heuristic tricks to select the most relevant/useful samples for GAN training are established in the literature, for example in \cite{Sinha2020}, although not for the specific case of reducing mode collapse.}

\clearpage
\section{\textblue{Additional Results}}
\label{sec:suppl_results}

\begin{figure*}[h]
\centering
\includegraphics[width=0.7\textwidth]{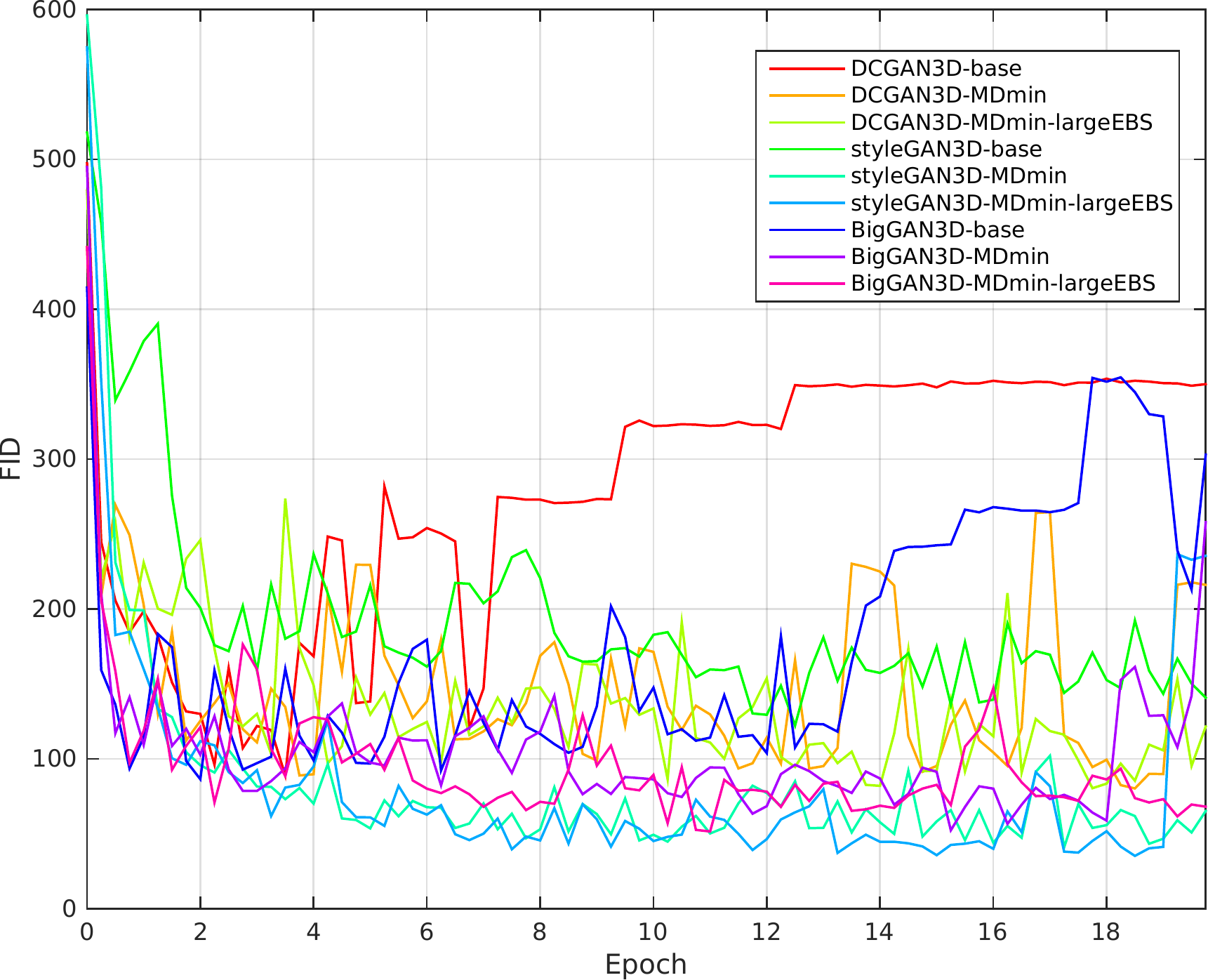}
\caption{\textblue{The progression of FID throughout training for the 9 comparative GAN models listed in Table \ref{table:comparitiveMethods_noLoss}, showing the run which provided the lowest FID score. }}
\label{fig:FID_graph}
\end{figure*}

\begin{figure*}[t]
\centering
\includegraphics[width=1\textwidth]{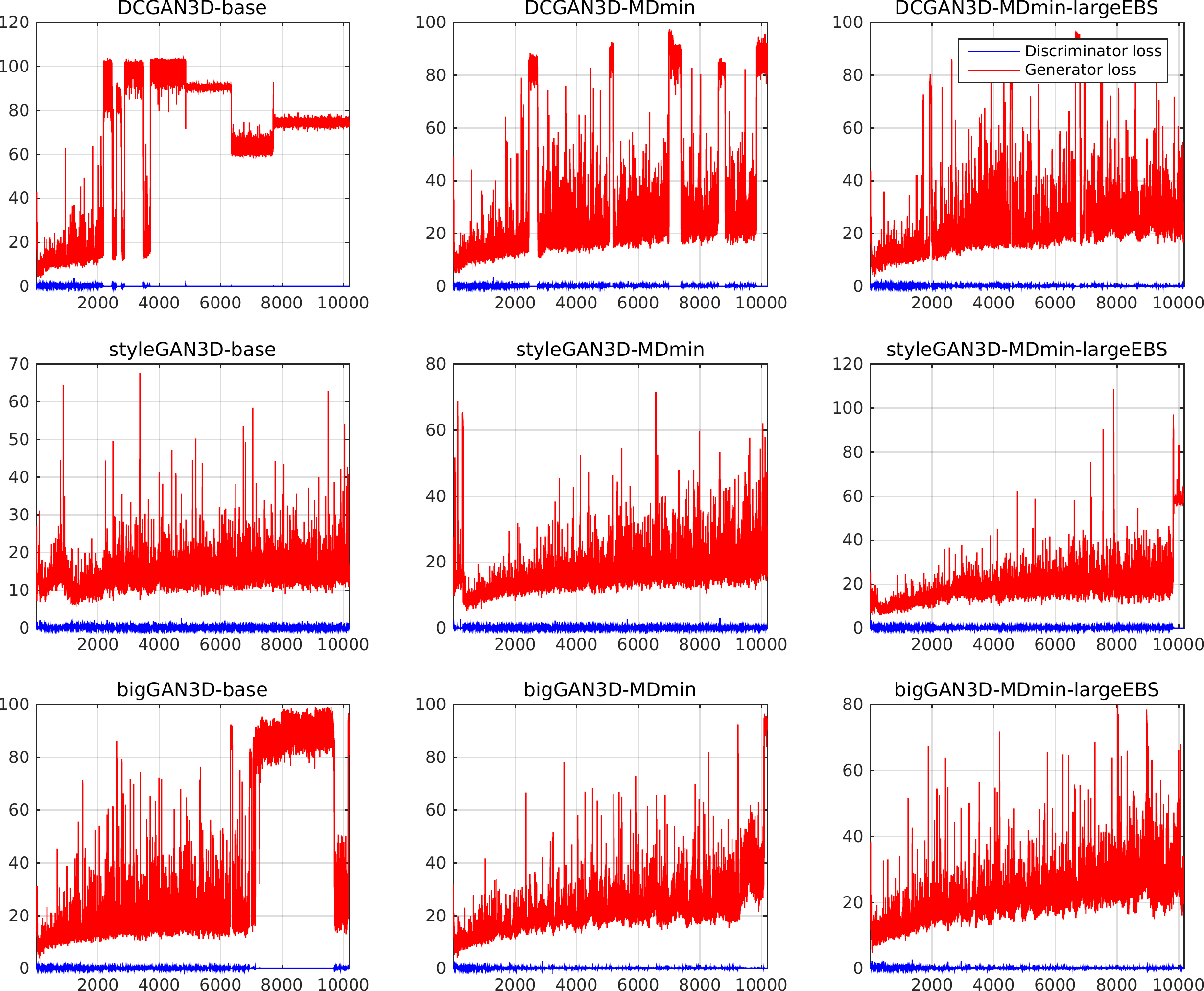}
\caption{\textblue{Training losses for the lowest-FID runs of each method in Table \ref{table:comparitiveMethods_noLoss}. Loss values are plotted every 14 steps, corresponding to the end of training with each CT image (see Section 3.3 for details.) }}
\label{fig:training_losses}
\end{figure*}

\begin{figure}[h]
\centering
\includegraphics[width=\textwidth]{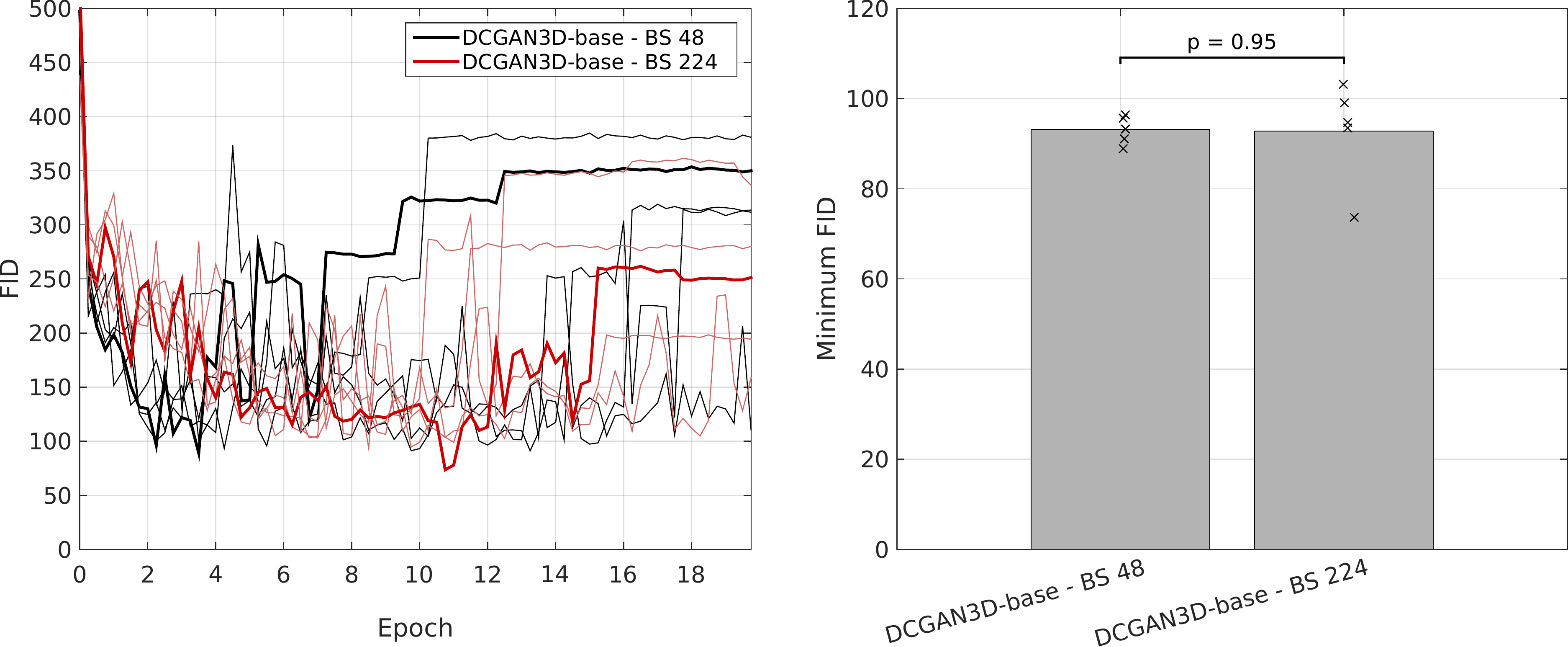}
\caption{\textblue{The effect of batch size on the DCGAN3D-base method. Increasing the batch size from 48 to 224 did not result in a significant improvement in terms of FID score, although on a single particular run the larger batch size did provide an FID of $73.7$.}}
\label{fig:ablation_DCGAN_BS}
\end{figure}


\begin{figure}[h]
\centering
\includegraphics[width=\textwidth]{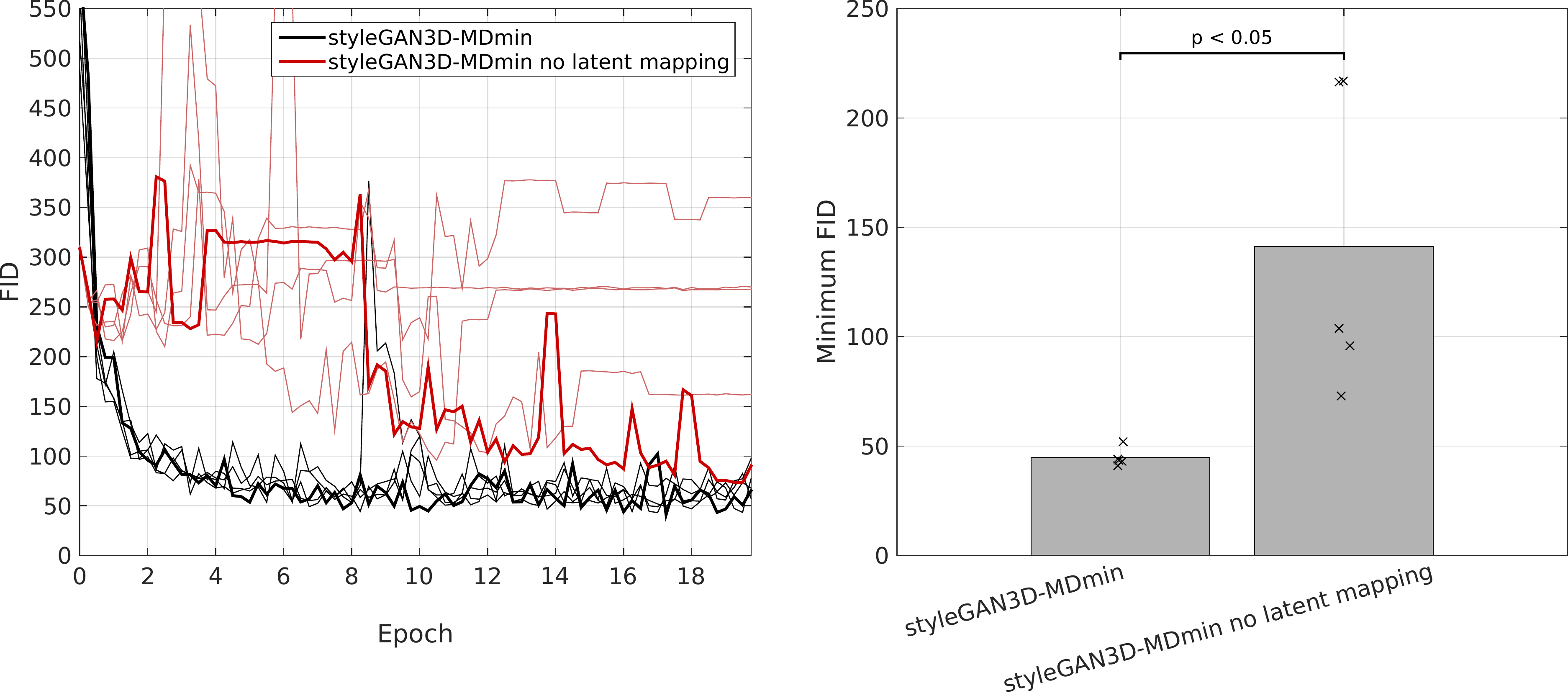}
\caption{\textblue{The effect of the latent mapping network on FID for the styleGAN3D-MDmin method. Removing the latent mapping network resulted in highly unstable training with a significant ($p<0.05$) increase in FID.}}
\label{fig:ablation_styleGAN_MDmin_nolatent}
\end{figure}


\end{document}